\documentclass{elsart}
\usepackage[latin1]{inputenc}
\usepackage[english]{babel}

\usepackage{amsmath}
\usepackage{amssymb}

\usepackage{graphicx}
\usepackage{hyperref}

\usepackage{bm}
\usepackage{graphicx}

\usepackage{dcolumn}

\begin{document}
\begin{frontmatter}

\title{Superconducting and excitonic quantum phase transitions in doped Dirac electronic systems}

\author[UFSJ]{Lizardo H. C. M. Nunes}
\ead{lizardonunes@ufsj.edu.br}
\author[UFSJ]{Ricardo L. S.  Farias}
and
\author[UFRJ]{Eduardo C. Marino}

\address[UFSJ]
{Departamento de Ci\^encias Naturais, Universidade Federal de S\~ao Jo\~ao del Rei, 36301-000 S\~ao Jo\~ao del Rei, MG, Brazil}

\address[UFRJ]
{Instituto de F\'{\i}sica, Universidade Federal do Rio de Janeiro, Caixa Postal 68528, Rio de Janeiro, RJ, 21941-972, Brazil}

\begin{abstract}
Material systems with Dirac electrons on a bipartite planar lattice and possessing superconducting and excitonic interactions are investigated both in the half-filling and doped regimes at zero temperature. 
%%%%%%%%%%%%
Excitonic pairing is the analog of chiral symmetry breaking of relativistic fermion theories and produces an insulating gap in the electronic spectrum. 
Condensed matter systems with such competing interactions display phenomena that are analogous to the onset of the chiral condensate and of color superconductivity in dense quark matter.
%%%%%%%%%%%%
Evaluation of the free-energy (effective potential) allows us to map the phases of the system for different values of the couplings of each interaction. 
%%%%%%%%%%%%
At half-filling, we show that Cooper pairs and excitons can coexist if the superconducting and excitonic interactions strengths are equal and above a quantum critical point, which is evaluated. 
%%%%%%%%%%%%
If one of the interactions is stronger than the other, then only the corresponding order parameter is non-vanishing and we do not have coexistence. 
%%%%%%%%%%%%
For nonzero values of chemical potential, the phase diagram for each interaction is obtained independently. Taking into account only the excitonic interaction, a critical chemical potential, as a function of the interaction strength, is obtained. 
%%%%%%%%%%%
If only the superconducting interaction is considered, the superconducting gap displays a characteristic dome as charge carriers are doped into the system and our results qualitatively reproduce the superconducting phase diagram of several compounds, like 122 pnictides and cuprate superconductors. 
%%%%%%%%%%%
We also analyze the possibility of coexistence between Cooper pairs and excitons and we show that, even if the excitonic interaction strength is greater than the superconducting interaction, as the chemical potential increases, superconductivity tends to suppress the excitonic order parameter. 
\end{abstract}

%%%%%%%%%%%%%%%%%%%%%%%%
%%%%%%%%%%%%%%%%%%%%%%%

\end{frontmatter}

\maketitle

\section{Introduction}
\label{int}

In the present paper, we investigate systems of Dirac electrons on a bipartite planar lattice possessing competing superconducting and excitonic interactions. 
Excitonic pairing is the analog of chiral symmetry breaking of relativistic fermion theories and produces an insulating gap in the electronic spectrum. Superconducting pairing, as we know, leads to zero resistance charge transport.
Condensed matter systems with such competing interactions display phenomena that
are analogous to the onset of the chiral condensate and of color superconductivity in dense quark matter, which is an active subject of investigation~\cite{Alford2008}.
The phase transitions obtained in such systems can be compared with the ones occurring in analog condensed matter systems
with Dirac fermions.

Indeed, several condensed matter systems
have been discovered in recent years whose active electrons have their kinematics
 governed by the Dirac equation, rather than by the Schr\"odinger
equation. Their dispersion relation, accordingly, has the same form as that of a
relativistic particle. The reason for this unusual behavior of electrons whose speed is
at least two orders of magnitude less than the speed of light can be ascribed to a
particular influence of the lattice background on the electronic properties.

One early example of such Dirac electrons in condensed matter can be found in
polyacetylene~\cite{poly1,poly2}.
The undoped system is in the half-filling regime. If we expand the one-dimensional
lattice tight-binding energy about the two Fermi points we obtain the linear dispersion
relation associated to a massless relativistic particle, the two components of the Dirac
field  corresponding to left and right moving electrons. Within a large temperature range,
any electronic excitations will have such a dispersion relation and consequently will obey
the Dirac equation.

In the high-Tc cuprate superconductors, Dirac points appear in the
intersection of the nodes of the $ d $-wave superconducting gap and the 2D-Fermi surface. Again,
the low-energy excitations will correspond exclusively to these points~\cite{cuprates}.

Quite recently, it has been experimentally found that the newest high-Tc superconductors,
namely the iron pnictides~\cite{pnic1,pnic2} also present electronic excitations whose properties
are governed by the Dirac equation. Theoretical results also support the existence of Dirac
electrons in the pnictides~\cite{Direlpnic,Direlpnic1}.

Another interesting example is that of graphene~\cite{Neto_RMP09},
In this case the tight-binding energy corresponding to an electron in the honeycomb lattice
presents a band structure such that the valence and conduction bands touch precisely in the vertices of two
inequivalent Dirac cones and any electronic excitations appearing in
the conduction band will have the dispersion relation of a relativistic massless particle
and their properties, accordingly, will be determined by the Dirac equation. In the
half-filling regime, with a completely filled valence band, we will have just a Fermi point
at the vertex, whereas in the presence of doping, a Fermi surface builds up.

For many technological applications, it would be interesting if graphene were a semiconductor, instead of a semimetal.
Therefore, the possibility of an insulator state or the appearance of an excitonic gap due to short range electronic interactions
in a system of massless Dirac fermions on the honeycomb lattice
has been the object of intense investigation~\cite{Araki2010,Armour2010,Weeks2010,Herbut2009,Liu2009}.
Nonetheless, early proposals that electron-electron interactions
could generate an electronic gap were investigated even before the
discovery of graphene~\cite{Khveshchenko,Gorbar2002},
where the gap opening is an analogue of the ``chiral symmetry'' breaking process
that occurs in massless quantum electrodynamics (QED) in two dimensions~\cite{ls,Appelquist1988}.
This is the type of excitonic instability that we consider in the present paper.

The fact that some of the above materials
become superconductors upon doping provides
a vast phenomenological motivation for the investigation
of the superconducting phase diagram of Dirac electrons systems.
Therefore, we analyse the effect of Cooper pair formation in such systems,
paying special attention to the cases where doping is
present, in which case we must introduce a finite chemical potential.

Indeed, we consider a quasi-two dimensional model consisting of Dirac electrons
both with superconducting and excitonic interactions.
In particular, we study the competition between these two interactions and
the resulting different situations that arise as we
independently vary the coupling parameter of each interaction term
and as we add charge carriers to the system.

The paper is organized as follows:
In Sections \ref{model} and \ref{T0} we obtain the continuum limit of our model
and we calculate the free-energy (effective potential) at zero temperature.
In Section \ref{mu=0}
we analyze  the phase diagram of the model in the absence of doping or impurities.
in Section \ref{muNeq0} the effect of doping is described by the introduction
of a non-vanishing chemical potential
and the physical effects of the interplay between the amount of doping
and the interactions strength bring interesting consequences,
which we fully analyze.
Each interaction is firstly taken independently and its respective phase diagram is obtained.
At the end of Section \ref{muNeq0} we combine our results and we investigate the possibility of coexistence
between Cooper pairs and excitons.

%%%%%%%%%%
\section{The model} \label{model}

Consider a system presenting a bipartite lattice formed of sublattices A and B, such as in graphene~\cite{Neto_RMP09} 
or in the pnictides~\cite{pnic2}.
Let
  $ a^{ \dagger }_{ i, \sigma } =
\sum_{ k }
e^{ i { \bf k } \cdot { \bf r}_{ i } }
\,
a^{ \dagger }_{ {\bf k }, \sigma } $ and $ b^{ \dagger }_{ i, \sigma} =
\sum_{ {\bf k } }
e^{ i { \bf k } \cdot { \bf r}_{ i }  }
\,
b^{ \dagger }_{ {\bf k }, \sigma}
$
be, respectively, electron creation operators, with spin $ \sigma $, on site $ i $  of sublattices $ A $ and  $ B $.
We take the non-interacting hamiltonian as
\cite{Neto_RMP09}
\begin{eqnarray}
H_{ t }
& =  &
- \mu \sum_{ {\bf k }, \sigma }
\left(
a^{ \dagger }_{ {\bf k }, \sigma } a_{ {\bf k }, \sigma }
+
b^{ \dagger }_{ {\bf k }, \sigma } b _{ {\bf k }, \sigma}
\right)
\nonumber \\
& &
-t \sum_{ {\bf k }, \sigma }
s_{ k }
\left(
a^{ \dagger }_{ {\bf k }, \sigma } b_{ {\bf k }, \sigma }
+ \mbox{h.c. }
\right)
\label{HGraphene}
\, ,
\end{eqnarray}
where the $ \mu $ is the chemical potential and the last line in the RHS of the above equation describes the hopping
of electrons between adjacent sites of different sublattices $ A $ and $ B $.

For the case of graphene, we would have
 $ t \approx 2.8 $ eV and, for the corresponding
honeycomb lattice, $ s_{ k } = 1
+ e^{ i { \bf k } \cdot { \bf  a }_{ 1 } }
+ e^{ i { \bf k } \cdot { \bf a }_{  2 } }
$
 where
$ {\bf a }_{ 1 } = a \hat{ e }_{ x } $
and
$
2 {\bf a }_{ 2 } = a \left(  \hat{ e }_{ x } - \sqrt{ 3 } \hat{ e }_{ y } \right) $,
with $ a $ as the lattice parameter.

We assume the interaction has two terms leading, respectively, to superconducting and excitonic instabilities.
The isotropic $ s $-wave pairing state is described by a BCS-type interaction as
\begin{eqnarray}
H_{ \mbox{\scriptsize{sc}} }
& = &
 - \lambda_{ \mbox{\scriptsize{sc}} }
\sum_{ {\bf k}, {\bf k}', \sigma}
\left(
a^{\dagger}_{ {\bf k }, \sigma }a^{\dagger}_{ - {\bf k }, -\sigma}
a_{ - {\bf k }', -\sigma }a_{ {\bf k }', \sigma }
\right.
\nonumber
\\
& &
\hspace{1.7cm}
+
\left.
b^{\dagger}_{ {\bf k }, \sigma }b^{\dagger}_{ - {\bf k }, -\sigma }
b_{ - {\bf k }', -\sigma }b_{ {\bf k }', \sigma }
\right)
\, ,
\label{EqHPairing}
\end{eqnarray}
 The value of $ \lambda_{ \mbox{\scriptsize{sc}} } $ is to be determined by some underlying microscopic theory,
which is not considered here. In the present paper, the strength of the superconducting interaction is a free
parameter of the hamiltonian and can take any arbitrary positive value.

The superconducting order parameter, by definition, is given by
\begin{equation}
 - \Delta
=
\lambda_{ \mbox{\scriptsize{sc}} }
\sum_{ \bf k }
\langle
a^{\dagger}_{ {\bf k }, \uparrow }a^{\dagger}_{ - {\bf k }, \downarrow }
\rangle
=
\lambda_{ \mbox{\scriptsize{sc}} }
\sum_{ \bf k }
\langle
b^{\dagger}_{ {\bf k }, \uparrow }b^{\dagger}_{ - {\bf k }, \downarrow }
\,
\rangle
\, .
\label{EqDefGap}
\end{equation}

Using a mean-field approximation, we insert the above expression in Eq.~Eq.~(\ref{EqHPairing}), thereby rewriting
the superconducting interaction as
    \begin{equation}
H^{ \mbox{\scriptsize{MF}} }_{  \mbox{\scriptsize{sc}} }
=
\sum_{ {\bf k } }
\Delta
\left(
a^{\dagger}_{ {\bf k }, \sigma }
a^{\dagger}_{ - {\bf k }, -\sigma}
+
b^{\dagger}_{ {\bf k }, \sigma }
b^{\dagger}_{ - {\bf k }, -\sigma }
\right)
+ \mbox{h.c. }
\; ,
\label{H_SC_MF}
\end{equation}
except for a constant term.

We define the four-component Dirac spinor as
$ \Psi_{ {\bf r }, \sigma } = \left( a_{ {\bf r }, \sigma }, b_{ {\bf r } , \sigma } \right) $ \cite{Neto_RMP09}.
In the spinor representation, the Coulomb interaction for an unitary volume reads
\begin{equation}
H_{  \mbox{\scriptsize{C}} }
=
\sum_{ \sigma, \sigma' }
\sum_{ {\bf r }, {\bf r}' }
\bar{ \Psi}_{ {\bf r }, \sigma }
\, \gamma^{ 0 } \,
\Psi_{ {\bf r }, \sigma }
%\left(
\, U( {\bf r }, {\bf r }' ) \,
%\right)
\bar{ \Psi}_{ {\bf r }', \sigma' }
\, \gamma^{ 0 } \,
\Psi_{ {\bf r }', \sigma '}
\; ,
\label{H_exc_MF}
\end{equation}
where $ \bar{ \Psi}_{ {\bf r }, \sigma } =  \Psi^{ \dagger }_{ {\bf r }, \sigma } \gamma^{ 0 } $,
with the Pauli matrix $ \gamma^{ 0 } $, and $ U( {\bf r }, {\bf r }' ) $ is the Coulomb potential.
A sufficiently strong Coulomb repulsion may set the stage for an excitonic instability which results in the opening of a gap in the electronic spectrum.  The transition is signaled by the development of the excitonic order parameter
\cite{Khveshchenko}
\begin{equation}
\sigma
=
\sum_{ \sigma }
\langle
\bar{ \Psi }_{ \sigma} \Psi_{ \sigma }
\rangle
=
\lambda_{ \mbox{\scriptsize{exc}} }
\sum_{ \sigma }
\langle
a^{ \dagger }_{ {\bf k }, \sigma  }
a_{ {\bf k }, \sigma }
-
b^{ \dagger }_{ {\bf k }, \sigma }
b_{ {\bf k }, \sigma }
\rangle
\, ,
\label{EqDefSigma}
\end{equation}
which is identified as a formal analog of the chiral symmetry breaking in relativistic fermion theories.
This is the type of excitonic instability that we focus upon below.
In momentum space, the corresponding excitonic pairing interaction in a mean-field approximation becomes,
\begin{equation}
H^{ \mbox{\scriptsize{MF}} }_{  \mbox{\scriptsize{exc}} }
=
\sum_{ {\bf k } , \sigma }
\sigma
\left(
a^{ \dagger }_{ {\bf k}  \sigma }
a_{ {\bf k } , \sigma }
-
b^{ \dagger }_{ {\bf k } , \sigma }
b_{ {\bf k } , \sigma }
\right)
\; .
%\label{H_exc_MF}
\end{equation}

Combining it with Eq.~Eq.~(\ref{HGraphene}) and Eq.~Eq.~(\ref{H_SC_MF}),
we obtain our model Hamiltonian, except for constant terms,
\begin{equation}
H =
\sum_{ {\bf k } }
\Phi^{ \dagger }_{ {\bf k } }
\mathcal{ A }_{ {\bf k } }
\Phi_{ {\bf k } }
\, ,
\label{H}
\,
\end{equation}
where the matrix $ \mathcal{ A }_{ \bf k } $ is given by
\begin{equation}
\mathcal{A}_{ {\bf k }  }
=
%\begin{pmatrix}
\left(
 \begin{array}{cccc}
\mu + \sigma       &    - t s_{ k }             &        0                       &        \Delta              \\
 - t s^{ * }_{ k }     &    \mu  - \sigma      &         \Delta               &        0                     \\
0                          &    \Delta^{ * }          &          \sigma - \mu   &        t s^{ * }_{ k }    \\
\Delta^{ * }           &     0                        &          t s_{ k }           &       - ( \mu - \sigma )  \\
\end{array}
\right)
%\end{pmatrix}
\label{EqMatrixA1}
\, ,
\end{equation}
    and
    $
    \Phi^{ \dagger }_{ \bf k }
=
a^{\dagger}_{ {\bf k }, \uparrow }
\;
b^{\dagger}_{ {\bf k }, \uparrow }
\;
b_{ -{\bf k }, \downarrow }
\;
a_{ -{\bf k }, \downarrow }
    $.

In the case of graphene, there are six Dirac points at the corners of the first Brillouin zone in the honeycomb
lattice; however, only two of them are non-equivalent. We denote them by $ {\bf K } =  - 4 \pi / 3 a \, \hat{ e }_{ x } $
and $ {\bf K }' =  4 \pi / 3 a \, \hat{ e }_{ x } $ and we expand Eq.~Eq.~(\ref{H}) in their vicinity. Collecting only
the linear terms for $ s_k $, and taking the continuum limit of our model, we obtain
 \begin{equation}
H^{ \mbox{\scriptsize{CL}} }
=
\sum_{ \alpha }
\int \frac{ d^{ 2 } k }{ \left( 2 \pi \right)^{ 2 } }
\,
\Phi^{ \dagger }_{ \alpha } ( k )
\, \mathcal{A}_{ \alpha }  \,
\Phi_{ \alpha } ( k )
- \frac{ | \Delta |^2 }{ \lambda_{ \mbox{\scriptsize{sc}} } }
- \frac{ \sigma^2 }{ \lambda_{ \mbox{\scriptsize{exc}} } }
\, ,
\label{EqHCL}
\end{equation}
    where $ \alpha = K $, $ K' $ denotes the Dirac points and the matrix $ \mathcal{A}_{ \alpha } $ is given by
Eq.~(\ref{EqMatrixA1}), replacing $ t s_k $ by $ - v_{ \rm{ F } } \left(  k_{ x } - i k_{ y } \right) $ or $ -
v_{ \rm{ F } } \left(  k_{ x } + i k_{ y } \right) $ in the vicinity of $ K  $ or $ K' $. Moreover, the Fermi velocity
is given by $ v_{ \rm{ F } } = \sqrt{ 3 }  t a / 2 $, with $ \hbar = 1$ from now on.

%%%%%%%%%%
\section{The effective potential} \label{T0}

The partition function in the complex time representation can be written as
\begin{eqnarray}
\mathcal{Z}
& = &
\frac{ 1 }{ \mathcal{ Z }_0 }
\int
\mathcal{D}\Pi
\,
\mathcal{D}\Phi
\nonumber \\
& &
\hspace{-0.3cm}
\times
\exp
\left[
\int_0^\beta d\tau
\int \frac{ d^2 k }{ ( 2 \pi )^2 }
\,
\left(
i \, \Pi \, \partial_{ \tau } \Phi
-
H[ \Pi, \Phi ]
\right)
\right]
\, ,
\label{Z}
\end{eqnarray}
where $ \Phi $ is an arbitrary set of fields, $ \Pi $ is their conjugate momentum and $ \mathcal{ Z }_0 $, the
vacuum functional.

Inserting Eq.~(\ref{EqHCL}) in the above expression, we can perform the gaussian integral over fermionic fields
to obtain an effective potential. Conversely, by integrating on $\Delta$ and $\sigma$ we would reobtain the original
fermionic quartic interaction hamiltonian by means of a Hubbard-Stratonovitch transformation.

The effective potential per Dirac point reads
\begin{eqnarray}
V_{ \mbox{\scriptsize{eff}} }
& = &
\frac{ | \Delta |^2 }{ \lambda_{ \mbox{\scriptsize{sc}} } }
+
\frac{ \sigma^2 }{ \lambda_{ \mbox{\scriptsize{exc}} } }
\nonumber
\\
& &
-
T
\sum_{ n = -\infty }^{ \infty }
\int \frac{ d^2 k }{ (2 \pi )^2 }
\nonumber \\
& &
\hspace{-0.2cm}
\times
\log
\left\{
\frac
{
\prod_{ j = 1 }^{ 4 }
\left[
i \omega_n - E_j
\right]
}
{
\prod_{ j = 1 }^{ 4 }
\left[
i \omega_n - E_j( \sigma = 0, \Delta = 0 )
\right]
}
\right\}
\label{Veff}
,
\end{eqnarray}
    where $ \omega_n =  ( 2 n + 1 ) \pi T $ is the Matsubara frequency for fermions (with the Boltzmann constant
$k_B = 1 $) and the four $ E_ j $'s are
   \begin{equation}
E_j
=
\pm
\sqrt{
{
| \Delta |^2 +
\left[
  \sqrt{ \sigma^2 + ( v_{ \rm{ F} } k)^2 }
\pm \mu
\right]^2
}
}
\label{Ej}
\, .
\end{equation}
Notice that Eq.~(\ref{Ej}) is the typical dispersion relation observed in effective models for quantum chromodynamics,
where $ E^{ - }_ j  $ corresponds to the energy required to create a particle or a hole above the Fermi surface
and $ E^{ + }_ j  $ is the corresponding antiparticle term \cite{Buballa2005}.
Moreover, the effective potential in Eq.~(\ref{Veff}) is the same obtained in \cite{Ratti2004},
where the formation of chiral and diquark condensates involving two quark flavours in QCD was investigated.
In their case, the values for the quark-antiquark and quark-quark interactions strengths were set equal by construction,
via a Fierz transformation applied to the Nambu-Jona-Lasinio model,
which is not the case in the present paper,
since $  \lambda_{ \mbox{\scriptsize{sc}} } $ and $ \lambda_{ \mbox{\scriptsize{exc}} } $
are taken to be free parameters.

At zero temperature, Eq.~(\ref{Veff}) becomes
\begin{eqnarray}
V_{ \mbox{\scriptsize{eff}} }
& = &
\frac{ | \Delta |^2 }{ \lambda_{ \mbox{\scriptsize{sc}} } }
+
\frac{  \sigma^2 }{ \lambda_{ \mbox{\scriptsize{exc}} } }
\nonumber \\
& &
\hspace{-0.95cm}
-
\int \frac{ d^2 k }{ (2 \pi )^2 }
\int_{ -\infty }^{\infty } \frac{ d\omega }{ 2 \pi }
\log
\left[
\frac
{
\gamma\left( \omega, k, \Delta, \sigma \right)
}{
\gamma\left( \omega, k, \Delta = 0, \sigma = 0 \right)
}
\right]
\label{Veff_T0}
\, ,
\end{eqnarray}
where
\begin{eqnarray}
\gamma\left( \omega, k, \Delta, \sigma \right)
& = &
\left(
| \Delta |^2 + \sigma^2 + v_{ \rm F }^2 k^2 + \omega^2
\right)^2 + \mu^4
\nonumber \\
& &
\hspace{-0.2cm}
+
2 \left( | \Delta |^2 - \sigma^2 - v_{ \rm F }^2 k^2 + \omega^2\right) \mu^2
\, .
\end{eqnarray}

Calculating the integral over $ \omega $ in Eq.~(\ref{Veff_T0}),
the effective
potential becomes,
\begin{eqnarray}
V_{ \mbox{\scriptsize{eff}} }
& = &
\frac{ | \Delta |^2 }{ \lambda_{ \mbox{\scriptsize{sc}} } }
+
\frac{ \sigma^2 }{ \lambda_{ \mbox{\scriptsize{exc}} } }
\nonumber \\
& &
-
\int \frac{ d^2 k }{ ( 2 \pi )^2}
\,
\left[
\sum_l \sqrt{ | \Delta |^2  + \xi^2_l } - 2 v_{ \rm F } k
\right]
\label{Veff_T0_1}
\, ,
\end{eqnarray}
where  $ \xi_l( x ) = \sqrt{ x + \sigma^2 } + l \mu $, with $ l = +1 $ or $ - 1$.
Introducing the momentum cutoff $ \Lambda / v_{ \rm F }$, we arrive at the following expression for the effective
potential,
\begin{eqnarray}
V_{ \mbox{\scriptsize{eff}} }
& = &
\frac{ | \Delta |^2 }{ \lambda_{ \mbox{\scriptsize{sc}} } }
+
\frac{ \sigma^2 }{ \lambda_{ \mbox{\scriptsize{exc}} } }
+
\frac{ 2 \Lambda^3 }{ 3 \alpha }
-
\frac{ 1 }{ 3 \alpha }
\sum_l
\left[
E_l^{ 3 }( \Lambda^2 )
-
E_l^{ 3 }( 0 )
\right]
\nonumber \\
& &
%-\hspace{-2cm}
+
\frac{ 1 }{ 2 \alpha }
\sum_l
%
%\left.
%
l \mu \,
\left\{
\,
\xi_l( x ) E_l( x )
\right.
\nonumber \\
& &
\left.
\left.
\hspace{1.6cm}
+
| \Delta |^2
\log\left[ \, \xi_l( x ) + E_l( x ) \, \right]
\,
\right\}
\right |_{ 0 }^{\Lambda^2 }
\label{Veff_T0_2}
\, ,
\end{eqnarray}
where $ \alpha  = 2 \pi v^2_{ \rm F }$ and $ E_l( x ) = \sqrt{ | \Delta |^2 + \xi_l^2( x ) } $.

In particular, for $ \mu = \sigma = 0 $, Eq.~(\ref{Veff_T0_2}) is reduced to
\begin{equation}
V_{ \mbox{\scriptsize{eff}} }
=
\frac{ | \Delta |^2 }{ \lambda_{ \mbox{\scriptsize{sc}} } }
-
\frac{ 2 }{ 3 \alpha }
\left[
\left(
| \Delta |^2 + \Lambda^2
\right)^{ \frac{ 3 }{ 2 } }
-
| \Delta |^3
-
\Lambda^3
\right]
\label{Veff_T0_s0_mu0}
\, ,
\end{equation}
which is exactly the same result obtained by some of us for a theory describing a single layer of Dirac electrons interacting
via a BCS-type superconducting interaction. Analysing the minima conditions for the above effective potential, we find
a quantum phase transition at the critical coupling $ \lambda_c = \alpha / \Lambda $~\cite{Marino2006} ,
leading to the onset of superconductivity in the system.

 In the remaining,  we analyze the conditions for the appearance of superconductivity or excitonic fluctuations at zero
temperature, taking into account the two competing interactions present in our model.

%%%%%%%%%%
\section{Zero chemical potential} \label{mu=0}

We analyze in this section the zero temperature phase diagram of our model in the absence of doping or impurities, namely,
at zero chemical potential.  We search for values of $ \Delta $ and $ \sigma $ that minimize the effective potential. The
conditions for the appearance of superconductivity or the excitonic fluctuations are furnished by the existence of nonzero
solutions of the respective order parameters, which minimize the effective potential.

The results are obtained in terms of an arbitrary energy cutoff $ \Lambda $ (momentum cutoff $ \Lambda / v_{ \rm F } $),
which is always provided by the lattice in condensed matter systems. Indeed, we have $ \Lambda = 2 \pi \hbar v_{ \rm F }/ a $.
Since $a$ is the smallest distance scale, $ \Lambda $ becomes a natural high-energy cutoff. This frequently happens in
condensed matter.  A familiar example in the case of conventional, phonon mediated superconductivity, is the Debye frequency
(energy)  a natural cutoff that emerges in BCS theory.

For $ \mu = 0 $, Eq.~(\ref{Veff_T0_2}) becomes
\begin{eqnarray}
V_{ \mbox{\scriptsize{eff}} }
& = &
-
\frac{ 2 }{ 3 \alpha }
\left[
\left(
| \Delta |^2 + \sigma^2 + \Lambda^2
\right)^{ \frac{ 3 }{ 2 } }
-
\left(
| \Delta |^2 + \sigma^2
\right)^{ \frac{ 3 }{ 2 } }
+
\Lambda^3
\right]
\nonumber \\
& &
+
\frac{ | \Delta |^2 }{ \lambda_{ \mbox{\scriptsize{sc}} } }
+
\frac{ \sigma^2 }{ \lambda_{ \mbox{\scriptsize{exc}} } }
\, ,
\label{Veff_T0_mu0}
\end{eqnarray}
which can be rewritten as
\begin{eqnarray}
V_{ \mbox{\scriptsize{eff}} }
& = &
\frac{ \Lambda^3 }{ \alpha }
\left\{
\frac{ \tilde{ \Delta }^2 }{ \tilde{ \lambda}_{ \mbox{\scriptsize{sc}} } }
+
\frac{ \tilde{ \sigma }^2 }{ \tilde{ \lambda}_{ \mbox{\scriptsize{exc}} } }
\right.
\nonumber \\
& &
\left.
\hspace{ -0.5 cm }
-
\frac{ 2 }{ 3 }
\left[
\left(
\tilde{ \Delta }^2 + \tilde{ \sigma }^2 + 1
\right)^{ \frac{ 3 }{ 2 } }
-
\left(
\tilde{ \Delta }^2 + \tilde{ \sigma }^2
\right)^{ \frac{ 3 }{ 2 } }
-
1
\right]
\right\}
\label{Veff_T0_mu02}
\, ,
\end{eqnarray}
where  $ \tilde{ \sigma } = \sigma / \Lambda^2 $ and  $ \tilde{ \Delta } =| \Delta | / \Lambda^2 $.

%%%%%%%%%%%%%%%%%%%%%%%%%%%%%%%%%%%%%%%%%%%%%%%%%%%%
\begin{figure}[ht]
\centerline
{
\includegraphics[clip, angle=-90, width=0.7\textwidth]{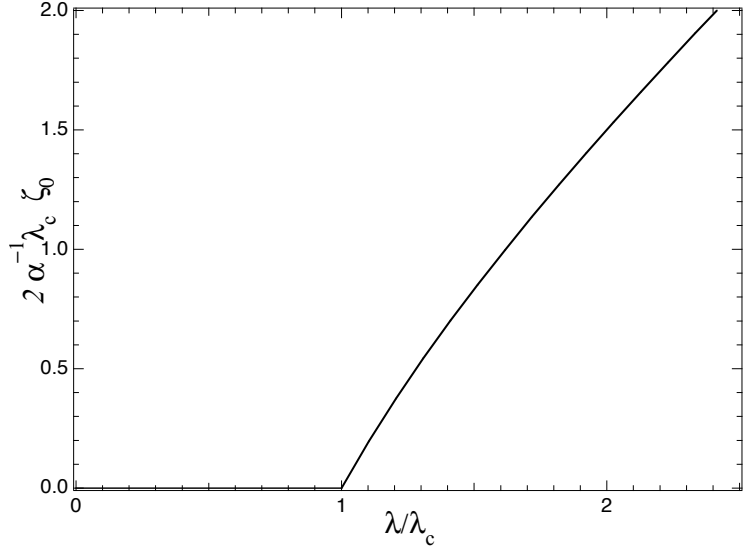}
}
\caption{ Order parameter $ \zeta_{0}$ as a function of the interaction strength
$ \lambda = \lambda_{ \mbox{\scriptsize{sc}} } = \lambda_{ \mbox{\scriptsize{exc}} }.$ }
\label{FigZeta}
\end{figure}
%%%%%%%%%%%%%%%%%%%%%%%%%%%%%%%%%%%%%%%%%%%%%%%%%%%%

First, let us assume the following situation:
$ \lambda = \lambda_{ \mbox{\scriptsize{sc}} } = \lambda_{ \mbox{\scriptsize{exc}} } $.
In such case, we can define a new order parameter,
$ { \zeta}^2  = |{ \Delta }|^2 + { \sigma }^2 $, which is inserted in Eq.~(\ref{Veff_T0_mu0}),
and we get exactly the same effective potential given by Eq.~(\ref{Veff_T0_s0_mu0}).
Therefore, analyzing the values of $ \zeta $ that minimize the effective potential, we arrive at
the following,
\begin{equation}
\zeta_0
= \sqrt{ |\Delta_0|^2 + \sigma^2_0 }
=
\left\{
\begin{array}{ l }
0 \hspace{2.cm}, \lambda < \lambda_c
\\
\frac{ \alpha \lambda }{ 2 }
\left(
\frac{ 1 }{ \lambda^2_c }
-
\frac{ 1 }{ \lambda^2 }
\right)
\hspace{0.cm},  \lambda \ge \lambda_c
\end{array}
\right.
\, .
\label{EqZeta0}
\end{equation}
where $ \zeta_0 $, $ \Delta_0 $ and $ \sigma_0 $ denote the solutions for the minimum of $ V_{ \mbox{\scriptsize{eff}} } $.
The plot of the order parameter as  a function of the interaction strength can be seen in Figure~\ref{FigZeta}. As in the
case when we had just a superconducting interaction and $ \mu = 0 $, we find a quantum phase transition at the critical
coupling $ \lambda_c = \alpha / \Lambda $. However, we now have also the possibility of coexistence between superconductivity
and the excitonic fluctuations if the interaction strengths are equal.

%%%%%%%%%%%%%%%%%%%%%%%%%%%%%%%%%%%%%%%%%%%%%%%%%%%%
\begin{figure}[ht]
\centering{
\includegraphics[clip, angle=0, width=0.7\textwidth]
{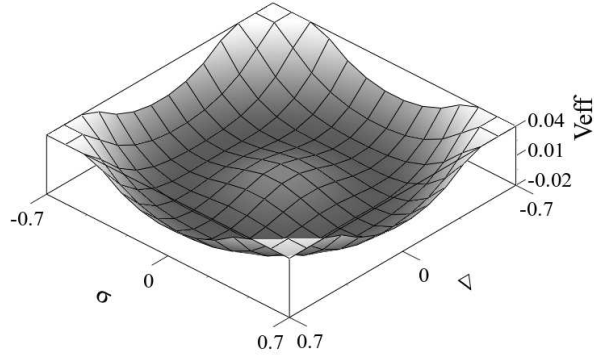}
}
\caption{ Effective potential including both cases: excitonic and superconducting
interactions. The parameters are taken as $\mu=0.0$, $ \tilde{ \lambda }_{ \mbox{\scriptsize{exc}} }=1.5$ and
$\tilde{ \lambda }_{ \mbox{\scriptsize{sc}} }=1.5$. All quantities are given in units of $\Lambda$.}
\label{FigVeff_mu0_lexcEqlsc}
\end{figure}
%%%%%%%%%%%%%%%%%%%%%%%%%%%%%%%%%%%%%%%%%%%%%%%%%%%%

Indeed, as can be seen in the tridimensional plot shown in Figure~\ref{FigVeff_mu0_lexcEqlsc},
the minima of the $ V_{ \mbox{\scriptsize{eff}} } $ presents a radial symmetry,
in agreement to the result in Eq.~(\ref{EqZeta0}).
Therefore, it is possible to find simultaneously nonzero values for $ \sigma $ and $ \Delta $ which minimise the effective
potential, indicating that excitons and Cooper pairs coexist in the system.

The same result was obtained in the framework of nuclear physics \cite{Ratti2004}, where the formation of chiral and diquark
condensates involving two quark flavors in QCD was investigated. In their case, the values for the quark-antiquark and
quark-quark interactions strengths were set equal by construction, via a Fierz transformation applied to the Nambu-Jona-Lasinio
model.

In our case, on the other hand, we are not restricted to the relation $ \lambda_{ \mbox{\scriptsize{sc}} } = \lambda_{
\mbox{\scriptsize{exc}} } $ and we shall analyze in what follows the situation for which $ \lambda_{ \mbox{\scriptsize{sc}} }
\neq \lambda_{ \mbox{\scriptsize{exc}} } $.

%%%%%%%%%%%%%%%%%%%%%%%%%%%%%%%%%%%%%%%%%%%%%%%%%%
\begin{figure}[ht]
\centering{
\includegraphics[angle=0, width=0.7\textwidth]{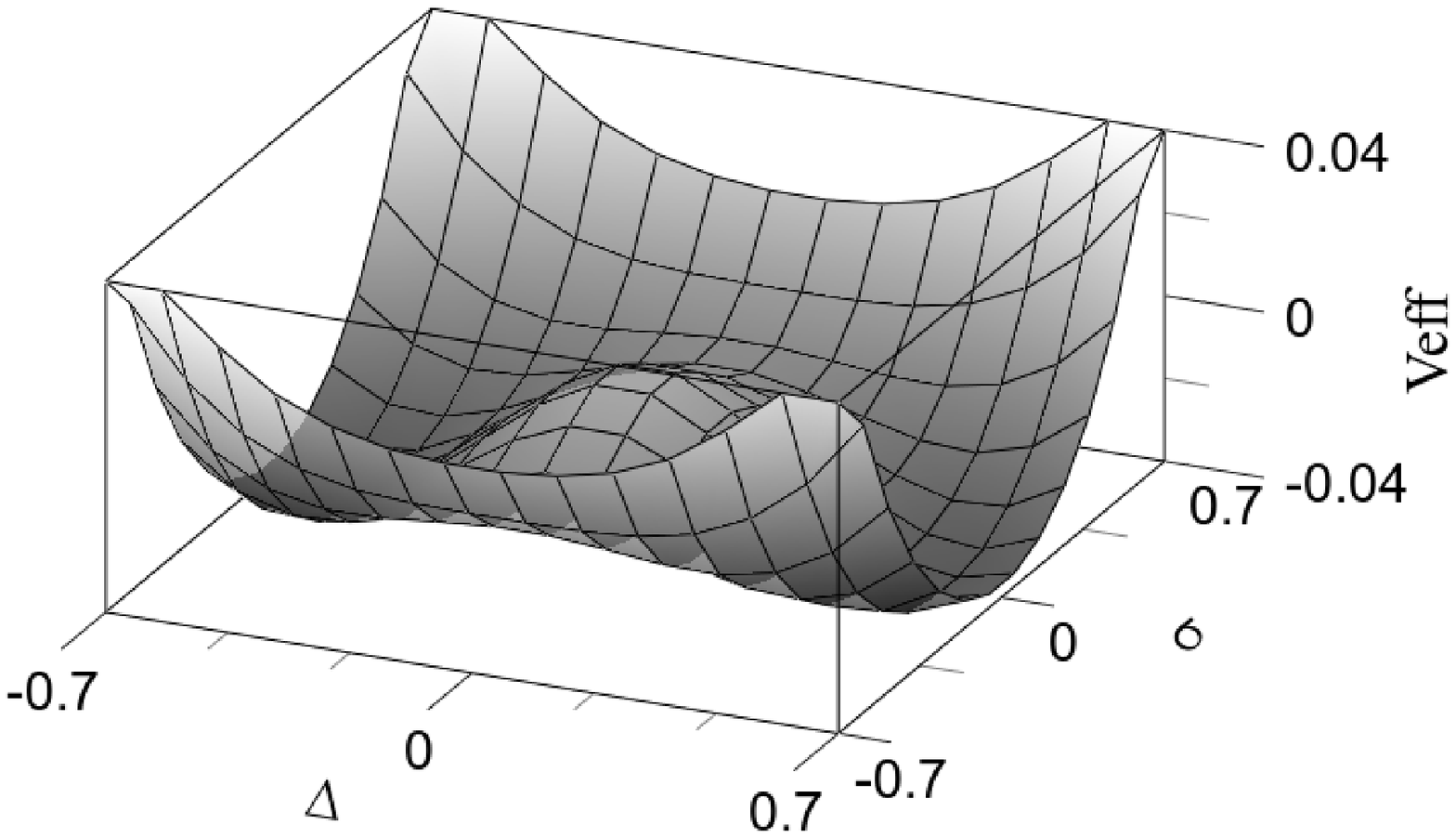}\\
\includegraphics[angle=0, width=0.7\textwidth]{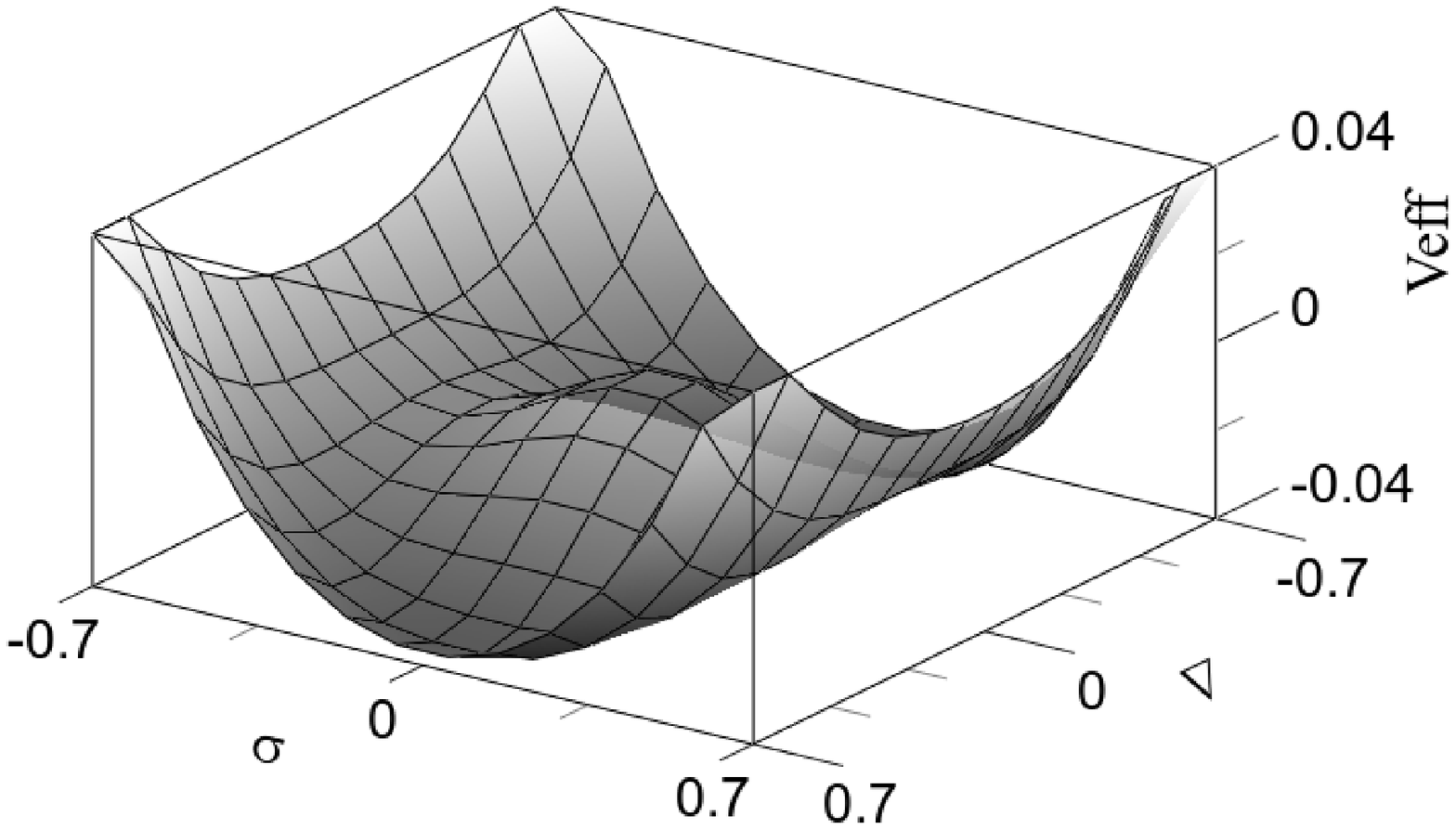}
        }
\caption{ Effective potential including both cases: excitonic and superconducting
interactions. The parameters are taken as $\mu=0.0$, $ \tilde{ \lambda }_{ \mbox{\scriptsize{exc}} }=1.5$ and
$ \tilde{ \lambda }_{ \mbox{\scriptsize{sc}} }=1.75$. All quantities are given in units of $ \Lambda $.}
\label{mu0_lsc_larger_lexc}
  \end{figure}
%%%%%%%%%%%%%%%%%%%%%%%%%%%%%%%%%%%%%%%%%%%%%%%%%%%%

We first assume that $ \lambda_{ \mbox{\scriptsize{sc}} } > \lambda_{ \mbox{\scriptsize{exc}} } $. Then, Eq.~(\ref{Veff_T0_mu02})
can be written as
\begin{eqnarray}
\tilde{ V }_{ \mbox{\scriptsize{eff}} }
\equiv
\frac{ \alpha }{ \Lambda^3 }
V_{ \mbox{\scriptsize{eff}} }
& = &
\frac{ \tilde{\zeta}^2 }{ \tilde{\lambda}_{ \mbox{\scriptsize{sc}} } } + \delta \tilde{\sigma}^2
\nonumber \\
& &
-
\frac{ 2 }{ 3 }
\left[
\left(
\tilde{ \zeta}^2 + 1
\right)^{ \frac{ 3 }{ 2 } }
-
\tilde{\zeta}^3
-
1
\right]
\label{Veff_T0_mu0_lexc.Neq.lsc}
\, ,
\end{eqnarray}
where
\begin{equation}
\delta =
\frac{
\tilde{\lambda}_{ \mbox{\scriptsize{sc}} } - \tilde{\lambda}_{ \mbox{\scriptsize{exc}} }
}{
\tilde{\lambda}_{ \mbox{\scriptsize{sc}} }  \tilde{\lambda}_{ \mbox{\scriptsize{exc}} }
}
> 0 \, ,
\label{Eqdelta}
\end{equation}
with
$  \tilde{ \lambda }_{ \mbox{\scriptsize{exc}} } = \lambda_{ \mbox{\scriptsize{exc}} } / \lambda_{ c } $
and
$  \tilde{ \lambda }_{ \mbox{\scriptsize{sc}} } = \lambda_{ \mbox{\scriptsize{sc}} } / \lambda_{ c } $.

Analysis of the second derivatives of the effective potential yields
the minimum of $ \tilde{ V }_{ \mbox{\scriptsize{eff}} } $ at the point
\begin{equation}
\left(  \Delta_0, \, \sigma_0 \right)
=
\left(
\frac{
\tilde{ \lambda }_{ \mbox{\scriptsize{sc} } }^2  - 1
}{
2 \tilde{ \lambda }_{ \mbox{\scriptsize{sc}} }
}
, \,
0
\right)
\, ,
\label{EqMinimum_lexc.LT.lsc}
\end{equation}
when
$ \lambda_{ \mbox{\scriptsize{sc}} } > 1 $
and $  \delta > 0 $
for positive values of $ \zeta_0 $ and $ \sigma_0 $.
Hence, $ \zeta_0 = \Delta_0 $
and we have only superconductivity in the system
provided $ \lambda_{ \mbox{\scriptsize{sc}} } > \lambda_{ \mbox{\scriptsize{exc}} } $.
Indeed, Figure~\ref{mu0_lsc_larger_lexc} shows the plot of $ V_{ \mbox{\scriptsize{eff}} } $ for
$ \lambda_{ \mbox{\scriptsize{sc}} } > \lambda_{ \mbox{\scriptsize{exc}} } $.
There is a single minimum for positive values of $ \Delta $ and $ \sigma $
exactly at the point given by Eq.~(\ref{EqMinimum_lexc.LT.lsc}).

%%%%%%%%%%%%%%%%%%%%%%%%%%%%%%%%%%%%%%%%%%%%%%%%%%%
\begin{figure}[ht]
    \centering{
        \includegraphics[angle=0, width=0.7\textwidth]{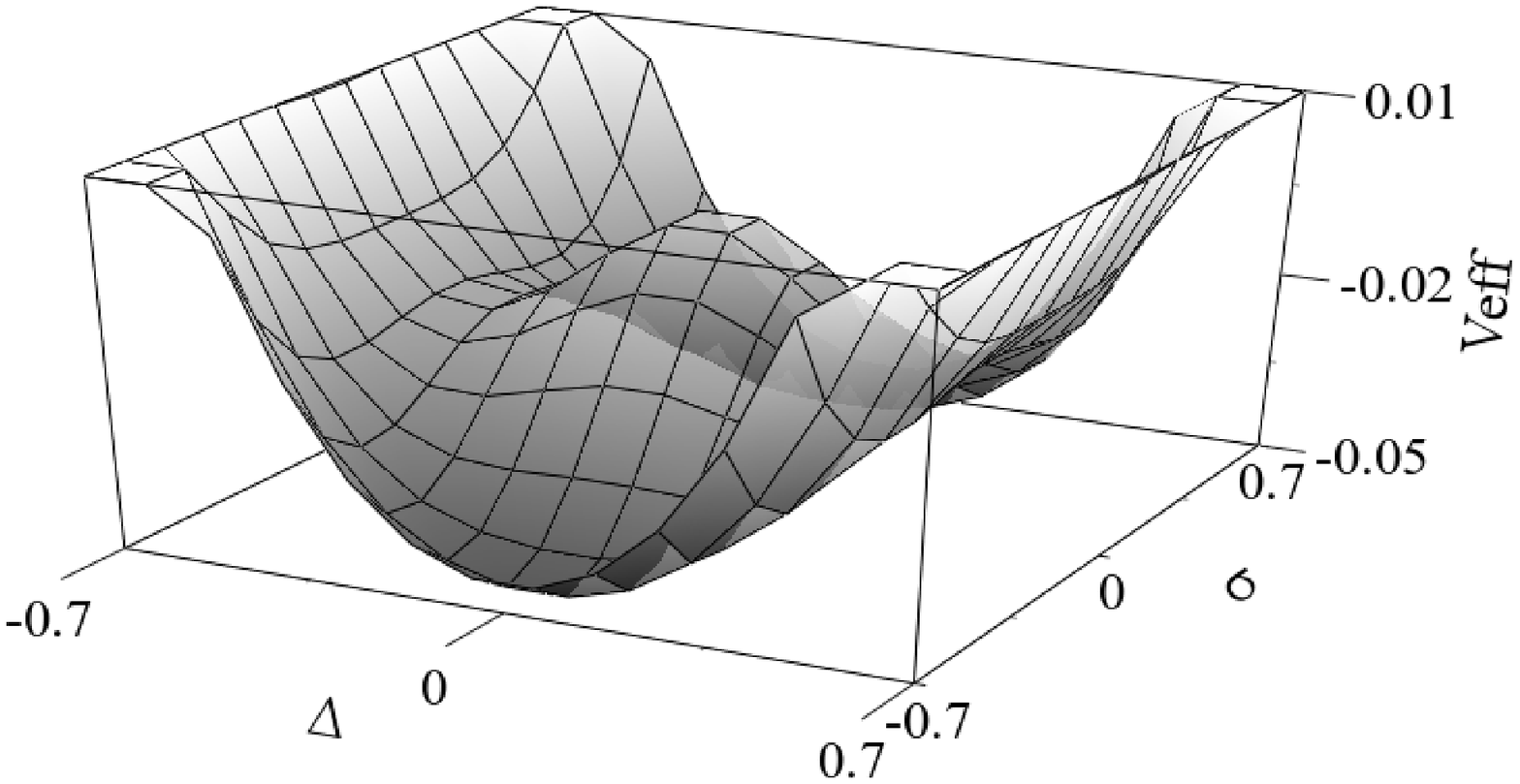}\\
        \includegraphics[angle=0, width=0.7\textwidth]{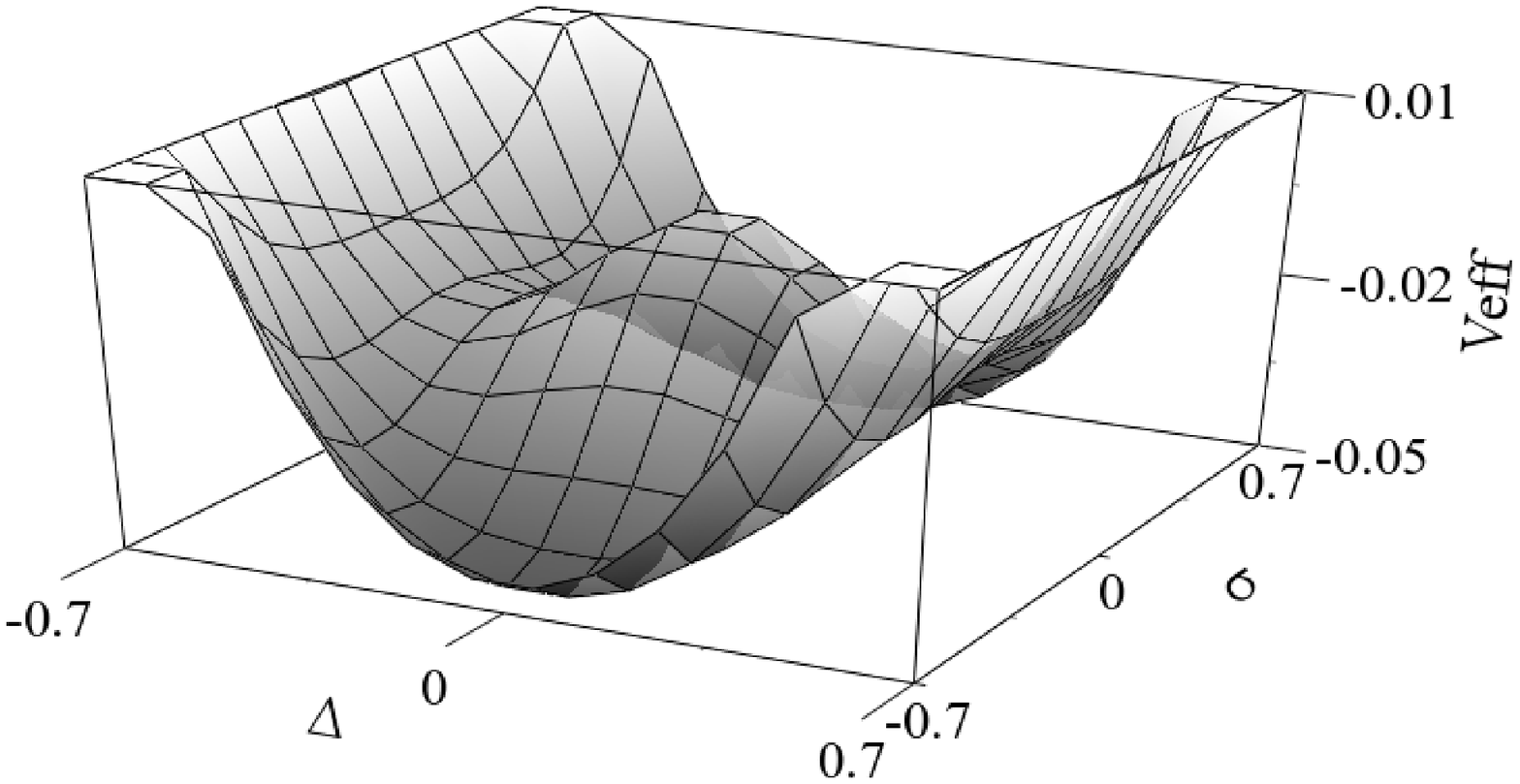}
}
\caption{ Plots of effective potential including both cases: excitonic and
superconducting interactions.These plots are the same but seen of different
points of view. Parameters are taken as $\mu=0.0$, $ \tilde{ \lambda }_{ \mbox{\scriptsize{exc}} }=1.75$ and
$ \tilde{ \lambda }_{ \mbox{\scriptsize{sc}} }=1.5$. All quantities are given in units of $\Lambda$.}
\label{mu0_lexc_larger_lsc}
\end{figure}
%%%%%%%%%%%%%%%%%%%%%%%%%%%%%%%%%%%%%%%%%%%%%%%%%%%

Accordingly, when $ \lambda_{ \mbox{\scriptsize{sc}} } < \lambda_{ \mbox{\scriptsize{exc}} } $,
the minimum becomes
\begin{equation}
\left(  \Delta_0, \, \sigma_0 \right)
=
\left(
0
, \,
\frac{
\tilde{ \lambda }_{ \mbox{\scriptsize{exc} } }^2  - 1
}{
2 \tilde{ \lambda }_{ \mbox{\scriptsize{exc}} }
}
\right)
\, ,
\label{EqMinimum_lexc.GT.lsc}
\end{equation}
when $ \tilde{ \lambda }_{ \mbox{\scriptsize{exc}} } > 1 $
and we only have excitons in the system.
Figure~\ref{mu0_lexc_larger_lsc} shows the plot of
$ V_{ \mbox{\scriptsize{eff}} } $
for $ \lambda_{ \mbox{\scriptsize{sc}} } < \lambda_{ \mbox{\scriptsize{exc}} } $
and the single minimum is given by Eq.~(\ref{EqMinimum_lexc.GT.lsc}).

In summary, in the undoped regime, where $ \mu = 0 $, the above results show that
we can have coexistence of excitons and Cooper pairs
whenever the corresponding interactions strengths are equal.
If one of the interactions is stronger than the other, then
only the corresponding order parameter is non-vanishing.

%%%%%%%%%%
\section{Finite chemical potential} \label{muNeq0}

Consider now the effective potential in Eq.~(\ref{Veff_T0_1}), in the
case of a nonzero chemical potential. Making the change of variable
$ x = ( v_{\rm F } k )^2 $, we get
\begin{eqnarray}
V_{ \mbox{\scriptsize{eff}} }
& = & \frac{ | \Delta |^2 }{ \lambda_{ \mbox{\scriptsize{sc}} } }
+
\frac{ \sigma^2 }{ \lambda_{ \mbox{\scriptsize{exc}} } }
+ \frac{ 2 \Lambda^3 }{ 3 \alpha }
\nonumber \\
& &
\hspace{-0.6cm}
-
\frac{ 1 }{ 2 \alpha }
\sum_l
\int_0^{ \Lambda^2 }
dx \,
\sqrt{ | \Delta |^2  + \left( \sqrt{ x + \sigma^2 } + l \mu \right)^2 }\, .
\label{Veff_T0_3}
\end{eqnarray}
We notice that $ V_{ { \mbox{\scriptsize{eff}} } }( \mu ) =  V_{ { \mbox{\scriptsize{eff}} }}( - \mu ) $. Therefore, given
$ \lambda_{ \mbox{\scriptsize{sc}} } $ and $ \lambda_{ \mbox{\scriptsize{exc}} } $, the minima for $ \mu > 0 $ are exactly
the same for $ \mu < 0 $ and we constrain our analysis to positive values of the chemical potential in this section.

In the following, we consider superconductivity and excitonic fluctuations separately, and then we combine our results to study
the superconducting and excitonic interactions together as the chemical potential increases.

%%%%%%%%%%%%%%%%%%%
\subsection{ $ \lambda_{ \mbox{\scriptsize{exc}} } \neq 0 $, $ \lambda_{ \mbox{\scriptsize{sc}} } = 0 $ } %\label{lexcNeq0}

In the absence of the superconducting interaction, the superconducting order parameter is zero and Eq.~(\ref{Veff_T0_3}) becomes
\begin{eqnarray}
\tilde{ V }_{ \mbox{\scriptsize{eff}} }
& \equiv &
\frac{ \alpha }{ \Lambda ^3 }
V_{ \mbox{\scriptsize{eff}} } ( \lambda_{ \mbox{\scriptsize{sc}} } = 0 )
\nonumber \\
& &
\hspace{-0.5cm}
=
\frac{ \tilde{ \sigma }^2 }{ \tilde{ \lambda }_{ \mbox{\scriptsize{exc}} } }
-
\frac{ 1 }{ 2 }
\sum_l
\int_0^{ 1 }
dy \,
\left|
\sqrt{ y + \tilde{ \sigma}^2 } + l \tilde{ \mu }
\right|
+ \frac{ 2 }{ 3 }
\, ,
\label{Veff_T0_s0}
\end{eqnarray}
where $  \tilde{ \lambda }_{ \mbox{\scriptsize{exc}} } = \lambda_{ \mbox{\scriptsize{exc}} } / \lambda_{ c } $
and the quantities within the modulus are divided by $ \Lambda $ and are denoted by $ \tilde{  \sigma } $ and $\tilde{ \mu } $,
since we have made the replacement $ y = x / \Lambda^2  $ in the above integral,
as in the previous section.

Since $ \mu > 0 $, we have that
$ \left| \sqrt{ y + \tilde{ \sigma}^2 } + \tilde{ \mu } \right| = \sqrt{ y + \tilde{ \sigma}^2 } + \tilde{ \mu } $,
for every positive values of $ \tilde{\sigma} $, $ \tilde{ \mu }$, $ y \in [ 0, 1 ] $ and $ l $
in the above equation.
For $ l = -1 $, on the other hand, according to the modulus definition, we have
\begin{equation}
\left|
\sqrt{ y + \tilde{ \sigma}^2 } - \tilde{ \mu }
\right|
=
\left\{
\begin{array}{ l }
%{*{20}c}
\sqrt{y + \tilde{ \sigma }^2 } - \tilde{ \mu }
\ \ \ \ \ \ \ ,
\tilde{ \mu } \le \sqrt{ y + \tilde{ \sigma }^2 }
\\
%\\
- \left(
\sqrt{y + \tilde{ \sigma }^2 } - \tilde{ \mu }
\right)
,
\tilde{ \mu } > \sqrt{ y + \tilde{ \sigma }^2 }
\end{array}
\right.
\, .
\label{System_T0_mu0}
\end{equation}
Hence, we have three different results for the effective potential:

\begin{itemize}
\item $ \tilde{ \mu } < \tilde{ \sigma } $:
\end{itemize}

In this case, the modulus becomes
\begin{equation}
\left|
\sqrt{ y + \tilde{ \sigma}^2 } - \tilde{ \mu }
\right|
=
\sqrt{y + \tilde{ \sigma }^2 } - \tilde{ \mu }
,
\;\;
\forall y \in [0,1]
\, ,
\label{EqMod1}
\end{equation}
and Eq.~(\ref{Veff_T0_s0}) yields
\begin{equation}
\tilde{ V }^{ ( 1 ) }_{ \mbox{\scriptsize{eff}} }
=
\frac{ \tilde{ \sigma }^2 }{ \tilde{\lambda}_{ \mbox{\scriptsize{exc}} } }
-
\frac{ 2 }{ 3 }
\left[
\left(
\tilde{ \sigma }^2 + 1
\right)^{ \frac{ 3 }{ 2 } }
-
\tilde{ \sigma }^3
-
1
\right]
\label{Veff_T0_s0_mod1}
\, ,
\end{equation}
which is exactly the same expression given by Eq.~(\ref{Veff_T0_s0_mu0})
if one replaces
$ \tilde{ \sigma } $ and $ \tilde{\lambda}_{ \mbox{\scriptsize{exc}} } $
for $ \zeta $ and $ \tilde{\lambda}_{ \mbox{\scriptsize{sc}} } $ respectively.

\begin{itemize}
\item $  \tilde{ \mu } > \sqrt{ 1 + \tilde{ \sigma }^2 } $:
\end{itemize}

The modulus becomes
\begin{equation}
\left|
\sqrt{ y + \tilde{ \sigma}^2 } - \tilde{ \mu }
\right|
=
-( \sqrt{y + \tilde{ \sigma }^2 } - \tilde{ \mu } )
,
\;\;
\forall y \in [0,1]
\, ,
\label{EqMod2}
\end{equation}
and Eq.~(\ref{Veff_T0_s0}) yields
\begin{equation}
\tilde{ V }^{ ( 2 ) }_{ \mbox{\scriptsize{eff}} }
=
\frac{ \tilde{ \sigma }^2 }{ \tilde{\lambda}_{ \mbox{\scriptsize{exc}} } }
-
\tilde{ \mu } + \frac{ 2 }{ 3 }
\label{Veff_T0_s0_mod2}
\, .
\end{equation}

\begin{itemize}
\item $ \tilde{ \sigma } < \tilde{ \mu } < \sqrt{ 1 + \tilde{ \sigma }^2 } $:
\end{itemize}

At this intermediary condition, Eq.~(\ref{Veff_T0_s0}) yields
\begin{eqnarray}
\tilde{ V }^{ ( 3 ) }_{ \mbox{\scriptsize{eff}} }
& = &
\frac{ \tilde{ \sigma }^2 }{ \tilde{ \lambda }_{ \mbox{\scriptsize{exc}} } }
+ \frac{ 2 }{ 3 }
-
\frac{ 1 }{ 2 }
\int_0^{ 1 }
dy \, \left(  \sqrt{ y + \tilde{ \sigma}^2 } + \tilde{ \mu } \right)
\nonumber \\
& &
-
\frac{ 1 }{ 2 }
\left[
\int_0^{ \mu^2 - \sigma^2 }
dy \, \left(  \sqrt{ y + \tilde{ \sigma}^2 } - \tilde{ \mu } \right)
\right.
\nonumber \\
& &
\hspace{1.cm}
\left.
\,
+
\int_{ \mu^2 - \sigma^2 }^{ 1 }
dy \, \left(  \sqrt{ y + \tilde{ \sigma}^2 } - \tilde{ \mu } \right)
\right]
\, .
\label{Veff_T0_s0_mod3}
\end{eqnarray}
and this equation  becomes
\begin{equation}
\tilde{ V }^{ ( 3 ) }_{ \mbox{\scriptsize{eff}} } =
\frac{ \tilde{ \sigma }^2 }{ \tilde{\lambda}_{ \mbox{\scriptsize{exc}} } }
-
\frac{ 2 }{ 3 }
\left[
\left(
\tilde{ \sigma }^2 + 1
\right)^{ \frac{ 3 }{ 2 } }
-
1
\right]
- \frac{ \mu^3 }{ 3 } + \mu \tilde{ \sigma }^2
\, .
\end{equation}

In the first case, the function $ \tilde{ V }^{ ( 1 ) }_{ \mbox{\scriptsize{eff}} } $ is minimized by
\begin{equation}
\tilde{ \sigma }_0
=
\left\{
\begin{array}{ l }
%{*{20}c}
0 \hspace{2.6cm}, \tilde{ \lambda }_{ \mbox{\scriptsize{exc} } } < 1
\\
\frac{1}{ 2 }
\left(
\tilde{ \lambda }_{ \mbox{\scriptsize{exc} } }
-
\tilde{ \lambda }^{ - 1 }_{ \mbox{\scriptsize{exc} } }
\right)
\hspace{0.2 cm}, \tilde{ \lambda }_{ \mbox{\scriptsize{exc} } } \ge 1
\end{array}
\right.
\, .
\label{EqSigma0_lsc0}
\end{equation}
Notice that this result does not depend of a given chemical potential, since the effective potential is $ \mu $-independent .
In the second case, $ \tilde{ V }^{ ( 2 ) }_{ \mbox{\scriptsize{eff}} } $
is minimized by $ \tilde{ \sigma } = 0 $.
In the third case, we do not find a nonzero minimum for $ \tilde{ V }^{ ( 3 ) }_{ \mbox{\scriptsize{eff}} } $ as well,
but only a maximum at the physical value of
$ \sqrt{ \left( \lambda^{ -1 } + \mu \right)^2 - 1  } $.
Therefore, the system does not present excitonic fluctuations when
$  \tilde{ \mu } > \sqrt{ 1 + \tilde{ \sigma }^2 } $ or $ \tilde{ \sigma }
< \tilde{ \mu } < \sqrt{ 1 + \tilde{ \sigma }^2 } $.

Combining all the above results, we can obtain the phase diagram of the system
for the excitonic fluctuations.

Let us start by the case $ \tilde{ \lambda }_{ \mbox{\scriptsize{exc} } } < 1 $
and $ \tilde{ \mu } = 0 $, in this case, $ \tilde{ \sigma}_0 = 0 $, according to
Eq.~(\ref{EqSigma0_lsc0}). Therefore, as the chemical potential increases, we
can only get the second or the third situations for the effective potential,
$ \tilde{ V }^{ ( 2 ) }_{ \mbox{\scriptsize{eff}} } $ or $ \tilde{ V }^{ ( 3 ) }_{
\mbox{\scriptsize{eff}} } $ respectively. Hence, we never reach a nonzero minimum for
the excitonic order parameter whenever $ \tilde{ \lambda }_{ \mbox{\scriptsize{exc} } }
< 1 $ for all values of $ \tilde{ \mu } $.

When $ \tilde{ \lambda }_{ \mbox{\scriptsize{exc} } } > 1 $, the analysis is subtler, but
also straightforward.  We start by the assumption that $ \tilde{ \mu } = 0 $. In this case,
we do have a nonzero minimum $ \tilde{ \sigma }_0 $ given by Eq.~(\ref{EqSigma0_lsc0}).
Now, if we take a small value for $ \tilde{ \mu } $, say $ \tilde{ \mu } = \bar{ \mu } $,
as $ \tilde{ \sigma } $ increases, the effective potential passes from the second situation,
$ \tilde{ V }^{ ( 2 ) }_{\mbox{\scriptsize{eff}} } $, to the first situation,
$ \tilde{ V }^{ ( 1 ) }_{ \mbox{\scriptsize{eff}} } $, at the point $ \tilde{ \sigma } =
\bar{ \mu } $ and the effective potential can only reach a minimum if $ \tilde{ \sigma }_0
> \bar{ \mu } $ as $ \tilde{ \sigma } $ continues to increase. However, even if
$ \tilde{ \sigma }_0 > \bar{ \mu } $, the effective potential may or may not possess a
minimum, because the minimum for $ \tilde{ V }^{ ( 1 ) }_{ \mbox{\scriptsize{eff}} } $ has to be
smaller than any value of $ \tilde{ V }^{ ( 2 ) }_{ \mbox{\scriptsize{eff}} } $, when $ \tilde{ \sigma } < \bar{ \mu } $.
Since  $ \tilde{ \lambda }_{ \mbox{\scriptsize{exc} } } > 1 $, we have that $ \tilde{ V }^{ ( 2 ) }_{ \mbox{\scriptsize{eff}} } $
is a monotonically increasing function for all $ \tilde{ \sigma }_0 < \bar{ \mu } $. Therefore, we get the following condition
for the appearance of a minimum for the effective potential: $ \tilde{ V }^{ ( 1 ) }_{ \mbox{\scriptsize{eff}} } ( \tilde{ \sigma }
= \tilde{ \sigma }_0 ) < \tilde{ V }^{ ( 2 ) }_{ \mbox{\scriptsize{eff}} } ( \tilde{ \sigma } = 0 ) $. So, whenever this
condition is satisfied, we have a nonzero excitonic order parameter,  $ \tilde{ \sigma }_0 =  \left( \tilde{ \lambda }^2 -
1 \right)/ 2 \tilde{ \lambda } $, for $ \tilde{ \lambda }_{ \mbox{\scriptsize{exc} } } > 1 $ and $ \tilde{ \mu } > 0 $.

Moreover, notice that the particular case when $ \tilde{ V }^{ ( 1 ) }_{ \mbox{\scriptsize{eff}} }
( \tilde{ \sigma } = \tilde{ \sigma }_0 ) < \tilde{ V }^{ ( 2 ) }_{ \mbox{\scriptsize{eff}} } ( \tilde{ \sigma } = 0 ) $ provides a critical
value for the chemical potential,
\begin{eqnarray}
\mu_c
& = &
\frac{ 1 }{ 2^{ \frac{ 2 }{ 3 } } }
\left[
-8 + \frac{ 3 }{ \tilde{\lambda}_{ \mbox{\scriptsize{exc}} } } - \frac{ 2 }{ \tilde{\lambda}_{ \mbox{\scriptsize{exc}} }^3 }
\right.
\nonumber \\
& &
\hspace{0.7cm}
+
\left.
%\hspace{0.6cm}
%+
\left(
2 + \frac{ 1 }{ \tilde{\lambda}_{ \mbox{\scriptsize{exc}} }^2 } + \tilde{\lambda}^2
\right)^{ \frac{ 3 }{ 2} }
- \tilde{ \lambda }_{ \mbox{\scriptsize{exc}} }^3
\right]^{ \frac{ 1 }{ 3 } }
\label{EqMuc}
\, .
\end{eqnarray}
As $ \tilde{ \mu } $ increases above the threshold $ \tilde{ \mu } >  \mu_c $, we can no longer find a minimum for the effective
potential. Hence, provided $ \tilde{ \lambda }_{ \mbox{\scriptsize{exc} } } > 1 $, $ \tilde{ \mu } =  \mu_c $ is also a quantum
critical point.

%%%%%%%%%%%%%%%%%%%%%%%%%%%%%%%%%%%%%%%%%%%%%%%%%%%
\begin{figure}[ht]
\centerline
{
\includegraphics[clip, angle=-90, width=0.7\textwidth]{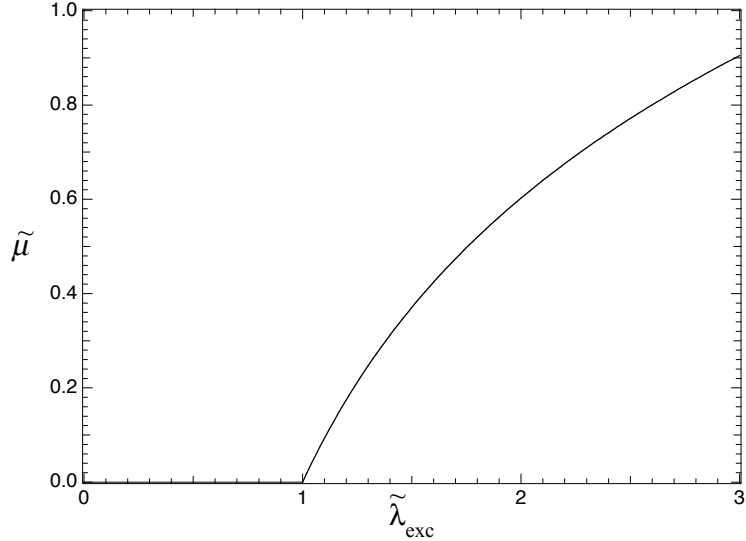}
}
\caption{ Phase diagram of the system taking into account only the excitonic interaction. The gray area indicates
the the presence of a nonzero excitonic order parameter in the system. The solid curve in the plot is given by the
critical chemical potential in Eq.~(\ref{EqMuc})}
\label{FigPD_lsc0}
\end{figure}
%%%%%%%%%%%%%%%%%%%%%%%%%%%%%%%%%%%%%%%%%%%%%%%%%%%
Figure~\ref{FigPD_lsc0} shows the phase diagram as a function of $ \tilde{ \lambda }_{ \mbox{\scriptsize{exc}} } $ and $ \tilde{ \mu } $.
The area below the curve for $ \mu_c   ( \tilde{ \lambda  }_{ \mbox{\scriptsize{exc}} } ) $ indicates the region where the system
presents excitonic fluctuations. To our best knowledge, this is the first time that this phase diagram was obtained.

%%%%%%%%%%%%%%%%%%%
\subsection{ $ \lambda_{ \mbox{\scriptsize{sc}} } \neq 0 $, $ \lambda_{ \mbox{\scriptsize{exc}} } = 0 $ } %\label{lexcNeq0}

In this section we analyze the phase diagram of the system taking into account only the presence of superconducting
interactions.
Hence, in the absence of excitonic interactions, Eq.~(\ref{Veff_T0_3}) becomes
\begin{eqnarray}
\tilde{ V }_{ \mbox{\scriptsize{eff}} }
& \equiv &
\frac{ \alpha }{ \Lambda ^3 }
V_{ \mbox{\scriptsize{eff}} }
( \lambda_{ \mbox{\scriptsize{exc}} } = 0 )
%\nonumber \\
%& = &
=
\frac{| \tilde{ \Delta }|^2 }{ \tilde{ \lambda }_{ \mbox{\scriptsize{sc}} } }
+ \frac{ 2 }{ 3 }
\nonumber \\
& &
-
\frac{ 1 }{ 2 \alpha }
\sum_l
\int_0^{ 1}
dy \,
\sqrt{| \tilde{ \Delta }|^2  + \left( \sqrt{ y } + l \tilde{ \mu } \right)^2 }
\, ,
\label{Veff_T0_D0}\end{eqnarray}
where the relevant quantities are divided by $ \Lambda $, as in the previous sections.
This problem has been previously studied by some of us for $ \mu = 0 $
\cite{Marino2006},
and some results for $ \mu \neq 0 $ have been previously reported considering the $ s $-wave and an exotic $ p $-wave pairing
\cite{Uchoa2007}.
For $ \mu = 0 $, the system undergoes a quantum phase transition at the critical coupling $ \lambda_{ \mbox{\scriptsize{sc}} }
= \alpha / \Lambda \equiv \lambda_c $
and the system becomes superconducting whenever $ \lambda_{ \mbox{\scriptsize{sc}} } > \lambda_c $.

Given values for $ \tilde{ \mu } $ and $  \tilde{ \lambda }_{ \mbox{\scriptsize{sc}} } $,
we can look for positive numerical values of $ \tilde{ \Delta } $ that minimise the effective potential,
given by Eq.~(\ref{Veff_T0_D0}). Once again, our analysis is constrained to positive values of $ \tilde{ \mu } $, since
$ \tilde{ V }_{ { \mbox{\scriptsize{eff}} } } ( \tilde{ \mu } ) =  \tilde{ V }_{ { \mbox{\scriptsize{eff}} } }
( - \tilde{ \mu } ) $.

%%%%%%%%%%%%%%%%%%%%%%%%%%%%%%%%%%%%%%%%%%%%%%%%%%%
\begin{figure}[ht]
\centerline
{
\includegraphics[clip, angle=-90, width=0.7\textwidth]{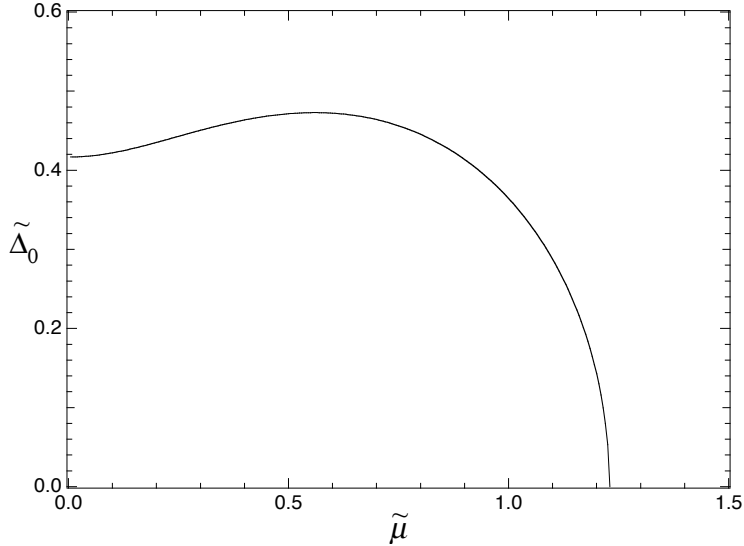}
}
\caption{ Plot of the superconducting gap $| \tilde{\Delta}_{0}| $ as a function of the chemical potential
$ \tilde{ \mu } $. The superconducting interaction is taken to be
$ \tilde{ \lambda}_{ \mbox{\scriptsize{sc}} } =1.5 $.
}
\label{FigD0vsmu_LscGT1}
\end{figure}
%%%%%%%%%%%%%%%%%%%%%%%%%%%%%%%%%%%%%%%%%%%%%%%%%%%

We start our discussion considering the case $ \tilde{ \lambda }_{ \mbox{\scriptsize{sc}} } > 1 $.
As can be seen in Figure~\ref{FigD0vsmu_LscGT1}, for $ \tilde{ \mu }= 0 $, the minimum of the effective potential
$ \tilde{ \Delta }_0 $ is given by Eq.~(\ref{EqZeta0}) implying $ \tilde{ \zeta }_0 \rightarrow \Delta_0 $. As
$ \tilde{ \mu } $ increases, $ \tilde{ \Delta }_0 $ reaches a maximum at an optimum value of the chemical potential
and $ \tilde{ \Delta }_0 $ decreases as $ \tilde{ \mu } $ increases even further.

%%%%%%%%%%%%%%%%%%%
%\begin{figure}[ht]
%\centerline{
%\includegraphics[clip, angle=0, width=0.49\textwidth]{./Figures/Delta0xmu_l08.png}
%}
%\caption{ Plot of the superconducting gap $ \tilde{\Delta}_{0} $ as a function of the chemical potential
%$ \tilde{ \mu } $. The superconducting interaction is taken to be
%$ \tilde{ \lambda}_{ \mbox{\scriptsize{sc}} } = 0.9 $. All values are in units of $\Lambda^{2}$.
%}
%\label{FigD0vsmu_LscLT1}
%\end{figure}
%%%%%%%%%%%%%%%%%%%

%%%%%%%%%%%%%%%%%%%%%%%%%%%%%%%%%%%%%%%%%%%%%%%%%%%
\begin{figure}[ht]
\centerline
{
\includegraphics[clip, angle=-90, width=0.7\textwidth]{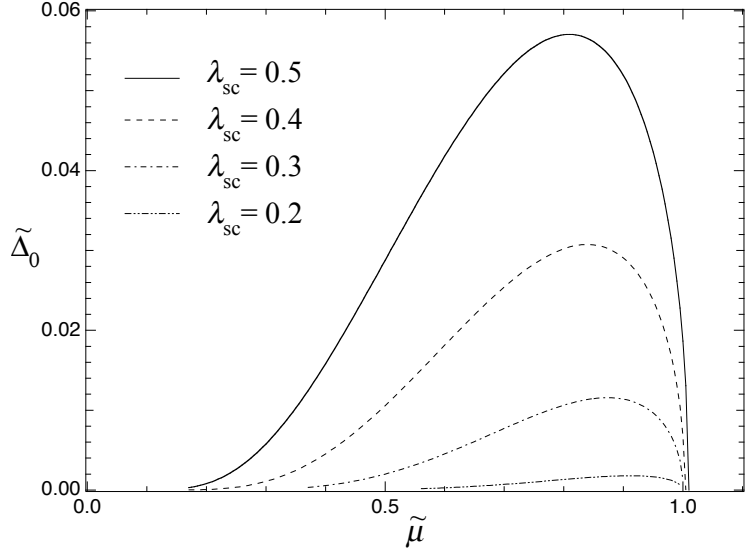}
}
\caption{ Plots of the superconducting gap
$ \tilde{\Delta}_{0} $ as a function of the chemical potential
$ \tilde{ \mu } $ for several values of $ \tilde{ \lambda}_{ \mbox{\scriptsize{sc}} } < 1 $.
}
\label{FigD0vsmu_LscLT1_2}
\end{figure}
%%%%%%%%%%%%%%%%%%%%%%%%%%%%%%%%%%%%%%%%%%%%%%%%%%%

For the case $ \tilde{ \lambda }_{ \mbox{\scriptsize{sc}} } < 1 $,
our results are shown in
Figure~\ref{FigD0vsmu_LscLT1_2}.
Starting at $ \tilde{ \mu } = 0 $, the system is in the normal state. However, as $ \tilde{ \mu } $ increases,
the system asymptotically becomes superconducting
and the order parameter also increases up to a maximum value at an optimal
chemical potential. As $ \tilde{ \mu } $ increases even further, $ \tilde{ \Delta }_0 $
decreases and the energy gap displays a dome-shaped plot.

Our results are consistent with \cite{Fukushima2007},
where chiral and diquark condensates are calculated for two-color and two-flavor QCD.
For a certain choice of parameters, the authors obtain the same dispersion relation given by Eq.~(\ref{Ej})
and their numerical results for $ \Delta $ also display a dome-shaped plot,
as can be seen in Fig. 1 of \cite{Fukushima2007}, for the choice of parameters I,
referred as the {\it weak-coupling case}.

Since the energy gap and the superconducting critical temperature $ T_{ c }$ are proportional,
our results qualitatively reproduce the superconducting phase diagram of several compounds, like 122 pnictides
and cuprate superconductors,
where the critical temperature displays a characteristic dome
as charge carriers are doped into the system.

A dome-like structure of the superconducting phase for strongly interacting two-dimensional Dirac fermions
has been previously obtained in \cite{Smith2009},
where the authors observed a dome-shaped plot of the superconducting phase at intermediate filling fractions,
surrounded by the normal phase for fillings close to unity or zero,
which is consistent to our results.

We conclude this section with a final remark: contrary to what happens at $ \mu = 0 $,
where superconductivity appears in the system only when
$  \lambda_{ \mbox{\scriptsize{sc}} } > \lambda_{c } $,
at finite chemical potential, we could always find a finite $ \tilde{ \mu } $
that provided nonzero superconducting gaps,
even at small values of $  \lambda_{ \mbox{\scriptsize{sc}} } $,
as can be seen in Figure~\ref{FigD0vsmu_LscLT1_2}, where
plots of $ \tilde{ \Delta }_{0} $ as a function of $ \tilde{ \mu } $
for small values of $  \tilde{ \lambda }_{ \mbox{\scriptsize{sc}} } $
are shown.
This result suggests that even for very small superconducting interaction strengths,
one can always find superconductivity as charge carriers are added to the system.

%%%%%%%%%%%%%%%%%%%
\subsection{ $ \lambda_{ \mbox{\scriptsize{sc}} } \neq 0 $, $ \lambda_{ \mbox{\scriptsize{sc}} } \neq 0 $ } %\label{lexcNeq0}

In the previous sections we have analysed the conditions for the appearance of the excitonic and superconducting order
parameters considering each interaction separately. In this section, we study the minima of the effective potential
taking into account both the interactions simultaneously and also the variation of the chemical potential.
From Eq.~(\ref{Veff_T0_3}), $ V_{ \mbox{\scriptsize{eff}} } $ can be written as
\begin{eqnarray}
\tilde{ V }_{ \mbox{\scriptsize{eff}} }
& \equiv &
\frac{ \alpha }{ \Lambda ^3 }
V_{ \mbox{\scriptsize{eff}} } ( \lambda_{ \mbox{\scriptsize{sc}} } \neq 0, \, \lambda_{ \mbox{\scriptsize{sc}} } \neq 0 )
\nonumber \\
& = &
\frac{| \tilde{ \Delta }|^2 }{ \tilde{ \lambda }_{ \mbox{\scriptsize{sc}} } }
+
\frac{ \tilde{ \sigma }^2 }{ \tilde{ \lambda }_{ \mbox{\scriptsize{exc}} } }
+ \frac{ 2 }{ 3  }
\nonumber \\
& &
\hspace{-0.3cm}
-
\frac{ 1 }{ 2 }
\sum_l
\int_0^{ \Lambda^2 }
dy \,
\sqrt{| \tilde{ \Delta }|^2  + \left( \sqrt{ y + \tilde{ \sigma }^2 } + l \tilde{ \mu } \right)^2 }
\,
\label{Veff_T0_4}
\end{eqnarray}
and, once again, we constrain ourselves to positive values of the chemical potential,
since $ V_{ \mbox{\scriptsize{eff}} } $ is even with respect to $ \mu $.

%%%%%%%%%%%%%%%%%%%%%%%%%%%%%%%%%%%%%%%%%%%%%%%%%%%
\begin{figure}[ht]
    \centering{
        \includegraphics[angle=0, width=0.7\textwidth]{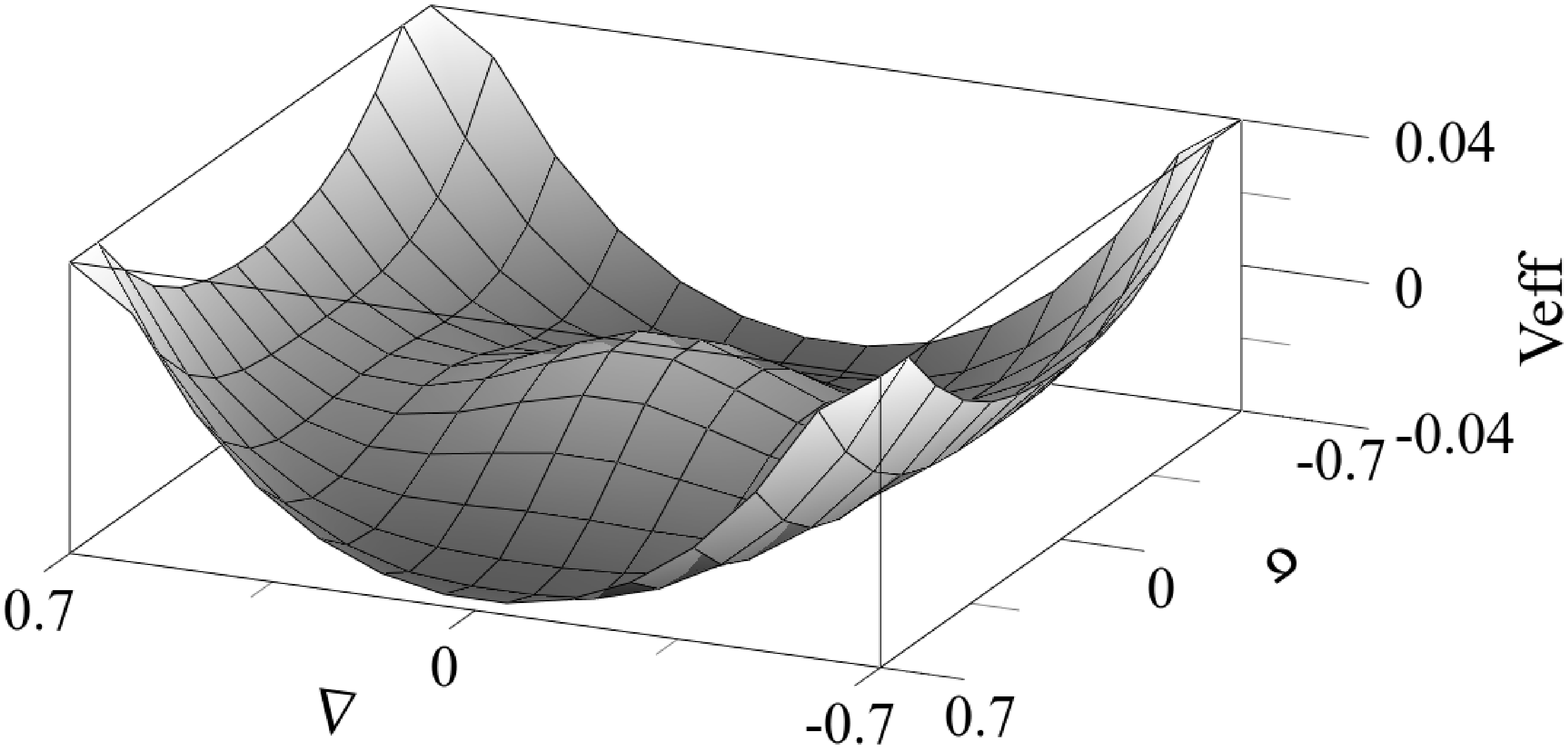}\\
	\includegraphics[angle=0, width=0.7\textwidth]{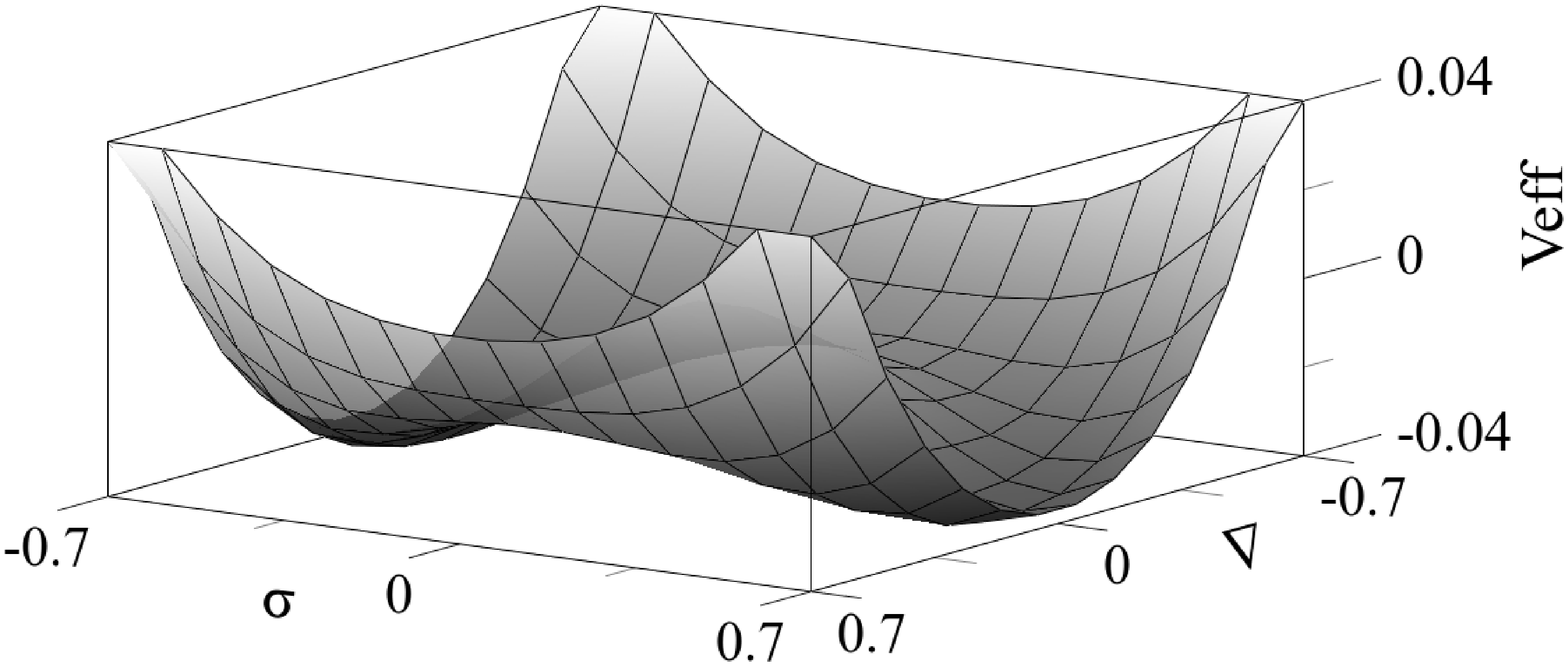}
}
\caption{ Effective potential including both cases: excitonic and superconducting
interactions.The parameters are taken as $ \tilde{ \mu }= 0.1 $, $ \tilde{ \lambda }_{ \mbox{\scriptsize{exc}} }=1.75 $ and
$ \tilde{ \lambda }_{ \mbox{\scriptsize{sc}} }=1.5 $. All quantities are given in units of $\Lambda $.}
\label{mu0p1_exc_larger_sc}
  \end{figure}
%%%%%%%%%%%%%%%%%%%%%%%%%%%%%%%%%%%%%%%%%%%%%%%%%%%

We start our analysis with $ \tilde{ \mu } = 0 $,
assuming that both $ \tilde{ \lambda}_{ \mbox{\scriptsize{exc}} }, \tilde{ \lambda }_{ \mbox{\scriptsize{sc}} } > 1 $.
We also impose that $ \tilde{ \lambda }_{ \mbox{\scriptsize{exc}} } $ is larger then $ \tilde{ \lambda }_{
\mbox{\scriptsize{sc}} } $. Therefore, as we have seen in Sec. \ref{mu=0}, since the interaction strengths are different
we cannot have coexistence of superconductivity and excitonic fluctuations. Actually, since $ \tilde{ \lambda }_{
\mbox{\scriptsize{exc}} } > \tilde{ \lambda }_{ \mbox{\scriptsize{sc}} } $, we have zero superconducting order
parameter and $ \tilde{ \sigma }_0 =  ( \tilde{ \lambda }^2_{ \mbox{\scriptsize{exc}} } - 1 )/ 2 \tilde{ \lambda }_{
\mbox{\scriptsize{exc}} } $. Indeed, as can be seen in Figure~\ref{mu0_lexc_larger_lsc}, the point $ \left( 0,
\tilde{\sigma}_0 \right) $ is the minimum of the effective potential for non-negative values of $ \tilde{\sigma}_{0} $
or $ \tilde{\Delta}_{0} $. Moreover, as we start to increase the values of the chemical potential, the minimum of
the effective potential remains the same as long as we constrain ourselves to small values of $ \tilde{ \mu } $, as
can be seen in Figure~\ref{mu0p1_exc_larger_sc}.

%%%%%%%%%%%%%%%%%%%%%%%%%%%%%%%%%%%%%%%%%%%%%%%%%%%
\begin{figure}[ht]
 \centering{\hspace{0.3cm}
	\includegraphics[clip, angle=0, width=0.7\textwidth]{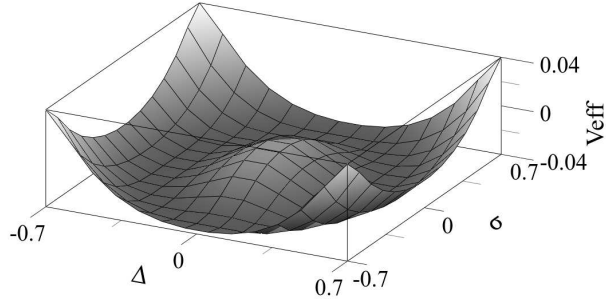}
}
\caption{ Effective potential including both cases: excitonic and superconducting
interactions.The parameters are taken as $ \tilde{ \mu }=0.3 $, $ \tilde{ \lambda }_{ \mbox{\scriptsize{exc}} }=1.75$ and
$ \tilde{ \lambda }_{ \mbox{\scriptsize{sc}} }=1.5$. All quantities are given in units of $\Lambda $.}
\label{mu0p3_exc_larger_sc}
  \end{figure}
%%%%%%%%%%%%%%%%%%%%%%%%%%%%%%%%%%%%%%%%%%%%%%%%%%%

Interesting results appear when  $ \tilde{ \mu } $ continues to increase,
since we regain the radial symmetry of the minima for the effective potential,
as can be seen in Figure~\ref{mu0p3_exc_larger_sc}.
In other words, we recover the coexistence of superconductivity and excitonic order fluctuations
even if $ \tilde{ \lambda }_{ \mbox{\scriptsize{exc}} } > \tilde{ \lambda }_{ \mbox{\scriptsize{sc}} } $
as the chemical potential reaches a certain value.

%%%%%%%%%%%%%%%%%%%%%%%%%%%%%%%%%%%%%%%%%%%%%%%%%%%
\begin{figure}[ht]
    \centering{
        \includegraphics[angle=0, width=0.7\textwidth]{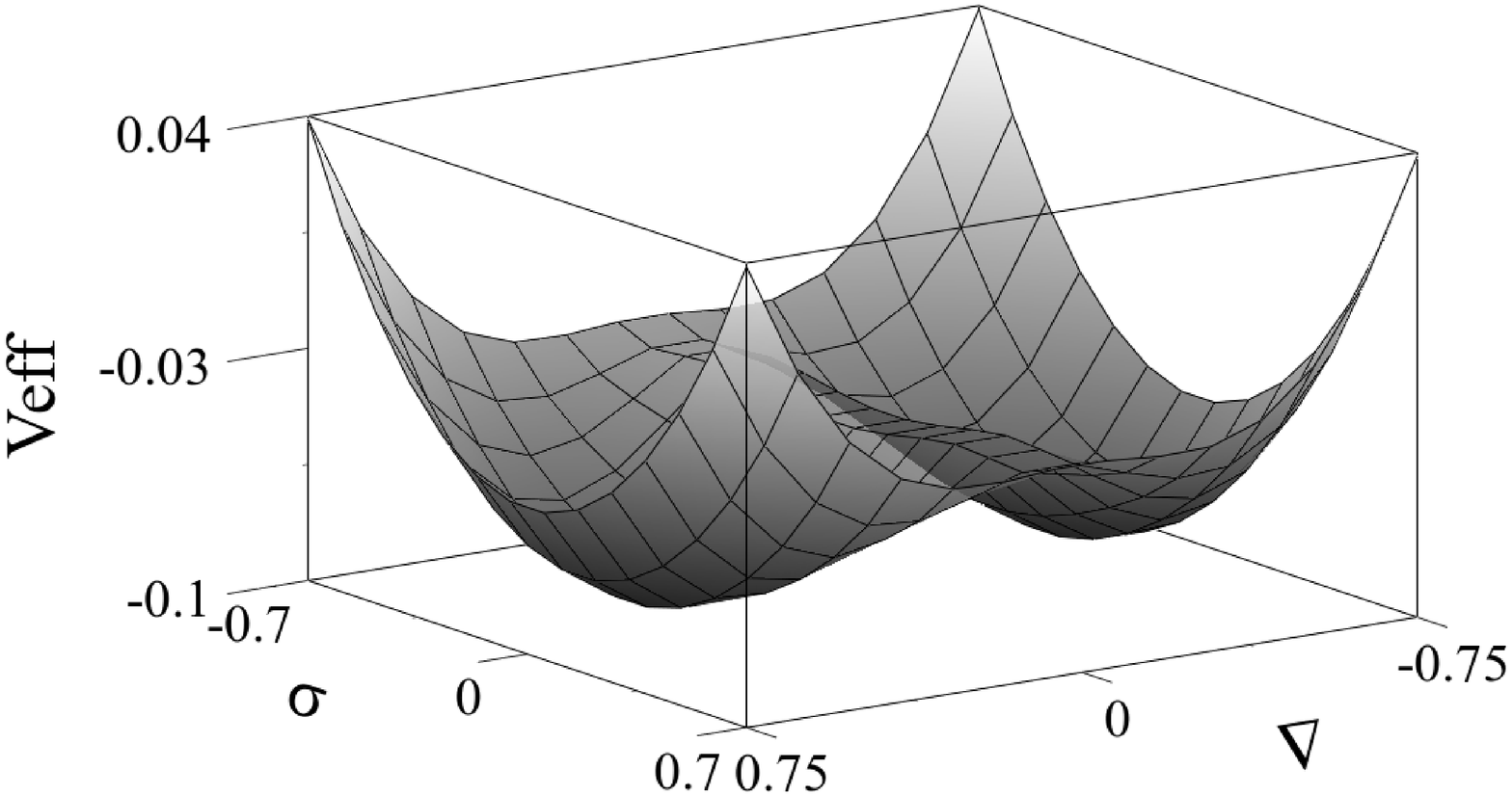}
        \includegraphics[angle=0, width=0.7\textwidth]{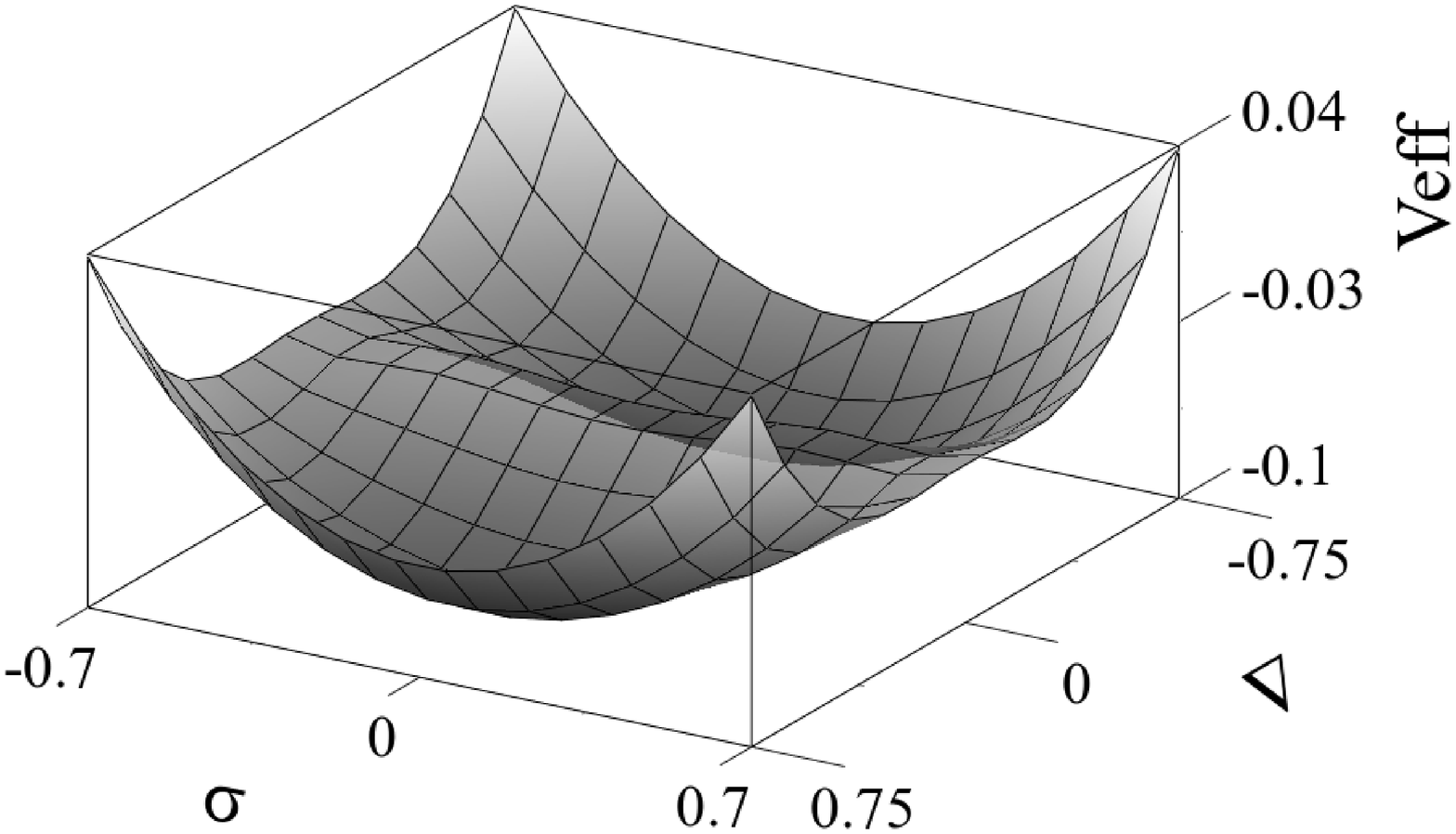}
}
\caption{ Effective potential including both cases: excitonic and superconducting
interactions.The parameters are taken as $ \tilde{ \mu } =0.5 $, $ \tilde{ \lambda }_{ \mbox{\scriptsize{exc}} }=1.75$ and
$ \tilde{ \lambda }_{ \mbox{\scriptsize{sc}} }=1.5$. All quantities are given in units of $ \Lambda $.}
\label{mu0p5_exc_larger_sc}
  \end{figure}
%%%%%%%%%%%%%%%%%%%%%%%%%%%%%%%%%%%%%%%%%%%%%%%%%%%

Furthermore, as the chemical potential increases even more, other interesting results show up,
as can be seen in in Figure~\ref{mu0p5_exc_larger_sc}.
In that case, the plot for the effective potential shows the minimum
at the point $ \tilde{ \sigma }_0 = 0 $ and  finite $ \tilde{ \Delta }_0 $ ,
which means that the excitonic fluctuations were suppressed from the system
and only superconductivity remains.
In other words, even if $  \tilde{ \lambda }_{ \mbox{\scriptsize{exc}} } > \tilde{ \lambda }_{ \mbox{\scriptsize{sc}} } $,
as charge carriers are added to the system,
doping effects eliminate the excitonic order parameter and favours superconductivity.

Our results are consistent to the results obtained in hadronic physics, where,
as the baryonic chemical potential increases, chiral symmetry in the system is restored
and color superconductivity sets in~\cite{Alford2008}.

%%%%%%%%%%%%%%%%%%%%%%%%%%%%%%%%%%%%%%%%%%%%%%%%%%%
\begin{figure}[ht]
    \centering{
        \includegraphics[angle=0, width=0.7\textwidth]{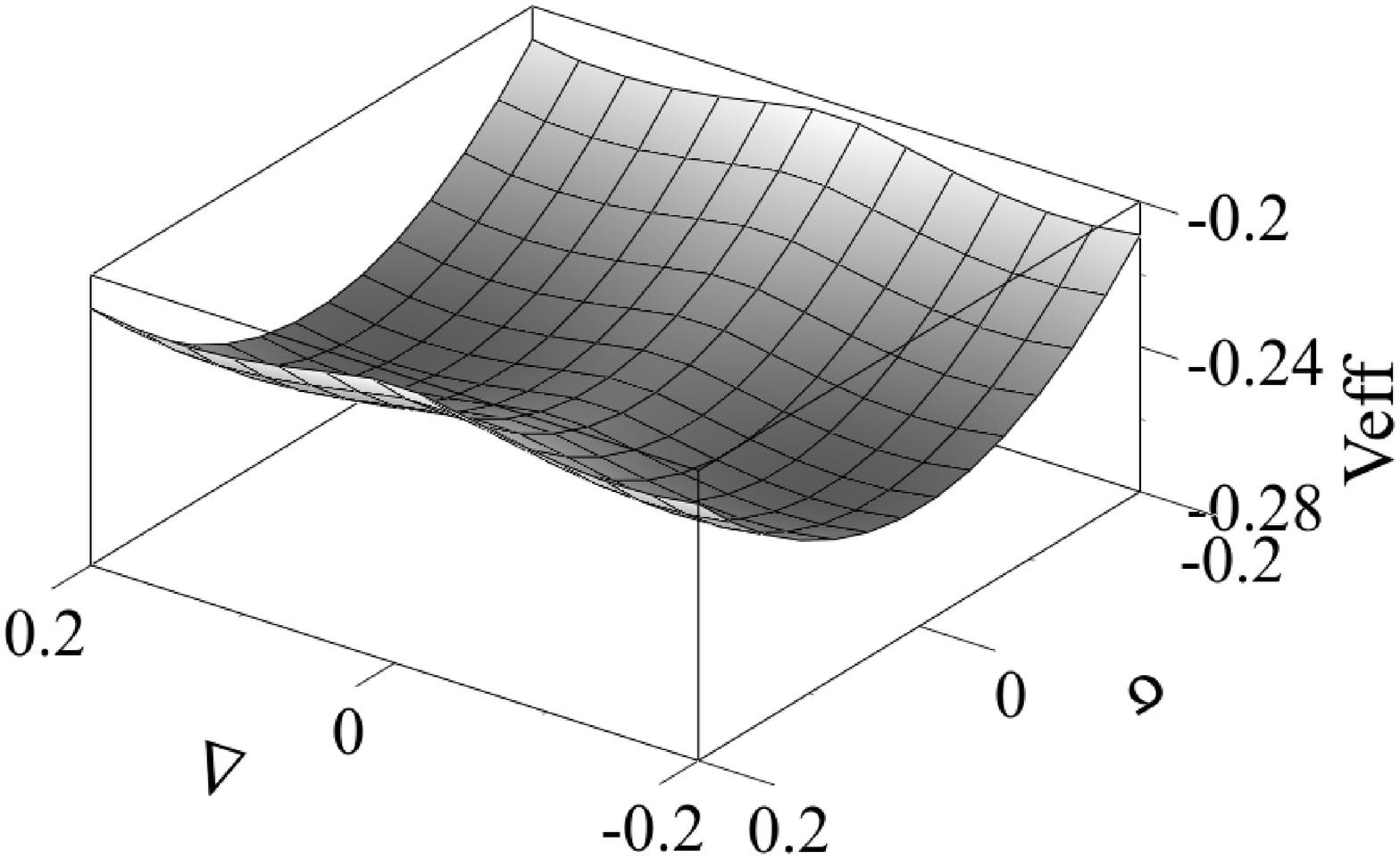}
        \includegraphics[angle=0, width=0.7\textwidth]{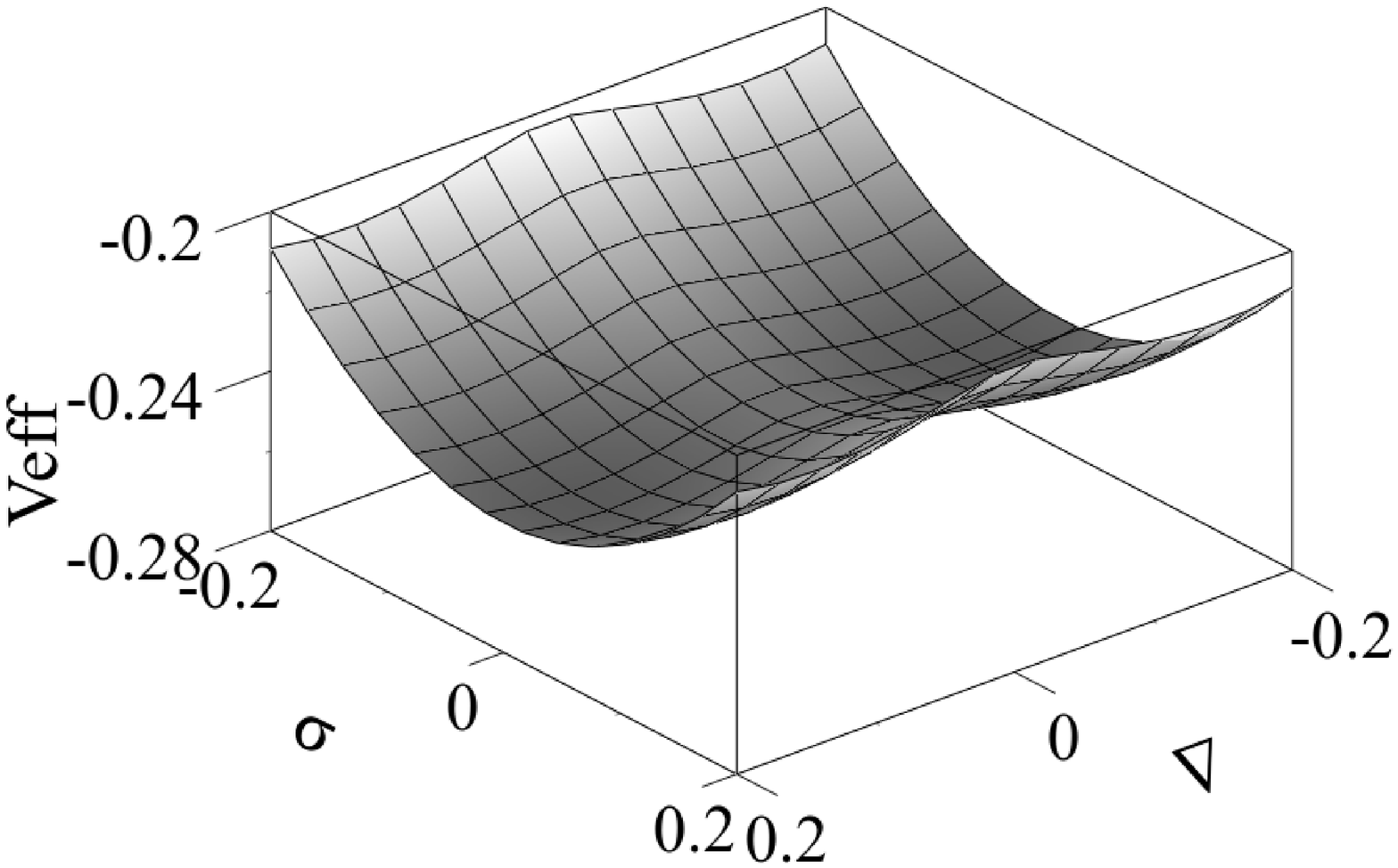}
}
\caption{ Effective potential including both cases: excitonic and superconducting
interactions.The parameters are taken as $ \tilde{ \mu } =0.5 $, $ \tilde{ \lambda }_{ \mbox{\scriptsize{exc}} } = 0.9 $ and
$ \tilde{ \lambda }_{ \mbox{\scriptsize{sc}} }=0.8$. All quantities are given in units of $ \Lambda $.}
\label{mu0p9_lexc0p9_l_sc0p8_viewDelta}
  \end{figure}
%%%%%%%%%%%%%%%%%%%%%%%%%%%%%%%%%%%%%%%%%%%%%%%%%%%

To conclude this section, we analyze the case when both
$ \tilde{ \lambda }_{ \mbox{\scriptsize{exc}} }, \tilde{ \lambda }_{ \mbox{\scriptsize{sc}} } < 1 $.
At zero value of the chemical potential,
we do not have the presence of superconductivity or excitonic fluctuations in the system.
However, as can be seen in Figure~\ref{mu0p9_lexc0p9_l_sc0p8_viewDelta},
the plot of effective potential show that we can find a nonzero superconducting gap
for,
$ \tilde{ \lambda }_{ \mbox{\scriptsize{exc}} }=0.9 $ and
$ \tilde{ \lambda }_{ \mbox{\scriptsize{sc}} }=0.8 $ at $ \tilde{ \mu } = 0.5 $
Therefore, even for small interactions with $ \tilde{ \lambda }_{ \mbox{\scriptsize{sc}} } <  \tilde{ \lambda }_{ \mbox{\scriptsize{exc}} } $,
there is the possibility of finding superconductivity
as charge carriers are added to the system.
These results indicate that,
as the chemical potential increases,
the insulating gap in the quasi-particle dispersion goes to zero,
and that superconductivity sets in.

\section{Conclusions}

Our investigation has produced several different results concerning 
different regimes of the coupling parameters of the superconducting and excitonic interaction terms on materials
possessing Dirac electrons. We also investigate the effects of doping on these.

At $ \mu = 0 $, we have shown that Cooper pairs and excitons
can coexist if the superconducting and excitonic interactions strengths are equal and above a quantum critical point,
$ \lambda = \lambda_{ \mbox{\scriptsize{sc}} } = \lambda_{ \mbox{\scriptsize{exc}} } > \alpha / \Lambda $.
If one of the interactions is stronger than the other,
then only the corresponding order parameter is non-vanishing
and we do not have coexistence.

For $ \mu \neq 0 $, taking into account only nonzero values for the excitonic interaction strength,
a critical chemical potential, as a function of $ \lambda_{ \mbox{\scriptsize{exc}} } $, was obtained,
as shown in  Eq.~(\ref{EqMuc}).
Given $ \lambda_{ \mbox{\scriptsize{exc}} } $,
values for the chemical potential below
$ \mu_c   ( \tilde{ \lambda  }_{ \mbox{\scriptsize{exc}} } ) $
indicate the presence of excitonic fluctuations in the system.
To our best knowledge, this is the first time that this phase diagram was obtained.

If there is only the superconducting interaction in the system,
a remarkable result was obtained for the case:
$ \lambda_{ \mbox{\scriptsize{sc}} } < \alpha / \Lambda $
the superconducting gap displays a dome-shaped curve as a function of the chemical potential,
indicating the appearance of superconductivity as charge carriers are added to the system. We emphasize that this
curve actually vanishes exponentially as $ \mu \rightarrow 0 $, hence superconductivity persists down to $ \mu = 0 $,
below some temperature. This is in agreement with the fact that Cooper's theorem must be valid for $ \mu \neq 0 $, when
a Fermi surface builds up.

Finally, we have analyzed the possibility of coexistence between Cooper pairs and excitons for $ \mu \neq 0 $ and we show that,
even if the excitonic interaction strength is greater than the superconducting interaction,
as the chemical potential increases, superconductivity tends to suppress the excitonic order parameter,
even if $  \tilde{ \lambda}_{ \mbox{\scriptsize{exc}} } > \tilde{ \lambda }_{ \mbox{\scriptsize{sc}} } $.

We have considered here exclusively the zero temperature situation. We are presently investigating
finite temperature effects on this system and will report the results elsewhere.

%%%%%%%%%%%%%%%%%%%%%
\ack
This work has been supported in part by CNPq, FAPEMIG and FAPERJ. We would like to thank H. C. G. Caldas and A. L. Mota for
discussions on related matters.

\end{document}